\documentclass[a4paper,fleqn,usenatbib]{mnras}

\usepackage{newtxtext,newtxmath}

\usepackage[T1]{fontenc}
\usepackage{ae,aecompl}

\usepackage{graphicx}	
\usepackage{amsmath}	
\usepackage{amssymb}	

\title[Orbital Period Change Across a Nova Eruption]{Precise Measures of Orbital Period, Before and After Nova Eruption for QZ Aurigae}

\author[B. E. Schaefer et al.]{Bradley E. Schaefer$^{1}$\thanks{E-mail: schaefer@lsu.edu},
David Boyd$^2$,
Geoffrey C. Clayton$^{1}$,
Juhan Frank$^1$
\newauthor
Christopher Johnson$^1$,
Jonathan Kemp$^3$,
Ashley Pagnotta$^4$,
Joseph O. Patterson$^5$, 
\newauthor
Miguel Rodr\'{i}guez Marco$^6$, \&
Limin Xiao$^{1}$
\\
$^{1}$Department of Physics and Astronomy, Louisiana State University, Baton Rouge, Louisiana, 70820, USA\\
$^2$BAA Variable Star Section, 5 Silver Lane, West Challow, Wantage, OX12 9TX, UK\\
$^3$Mittelman Observatory, Middlebury College, Middlebury, Vermont 05753, USA\\
$^4$Department of Physics and Astronomy, College of Charleston,  66 George Street, Charleston, South Carolina, 29424 USA\\
$^5$Department of Astronomy, Columbia University, New York City, New York, 10027, USA\\
$^6$AAVSO, Alberdi, 42, 2F, 28029 Madrid, Spain\\
}

\date{Accepted XXX. Received YYY; in original form ZZZ}

\pubyear{2018}

\begin{document}
\label{firstpage}
\pagerange{\pageref{firstpage}--\pageref{lastpage}}
\maketitle

\begin{abstract}
For the ordinary classical nova QZ Aurigae (which erupted in 1964), we report 1317 magnitudes from 1912--2016, including four eclipses detected on archival photographic plates from long before the eruption.  We have accurate and robust measures of the orbital period both pre-eruption and post-eruption, and we find that the orbital period {\it decreased}, with a fractional change of -290.71$\pm$0.28 parts-per-million across the eruption, with the orbit necessarily getting smaller.  Further, we find that the light curve outside of eclipses and eruption is flat at near B=17.14 from 1912--1981, whereupon the average light curve starts fading down to B=17.49 with large variability.  QZ Aur is a robust counter-example against the Hibernation model for the evolution of cataclysmic variables, where the model requires that all novae have their period increase across eruptions.  Large period decreases across eruptions can easily arise from mass imbalances in the ejecta, as are commonly seen in asymmetric nova shells.  
\end{abstract}




\section{Introduction}

During the thermonuclear eruption of a classical nova (CN), the white dwarf in the close interacting binary system ejects $\sim$10$^{-5}$ M$_{\odot}$, so by Kepler's Law, the orbital period, $P$, should change by getting larger.  Such a period change across the nova event, $\Delta P$=$P_{post}$-$P_{pre}$, must make the two stars separate slightly, a possibly substantial drop in the accretion rate and in the brightness of the binary system.  Indeed, the $\Delta P$$>$ 0 case can be used as the mechanism to drive a cycle of evolution for cataclysmic variables (CVs).  The `Hibernation model' for CV evolution (Shara et al. 1986) has this basic mechanism driving a post-eruption nova system to have a substantial drop in accretion and brightness in the decades and century after the nova, transforming the CV successively into systems of lower accretion rates (Z Cam stars, then dwarf novae, then disconnected and quiescent).  After a long time interval during which the system is drawn together by angular momentum loss from the emission of gravitational waves, the binary starts up accretion again, increasing until the accretion rate is high, when another nova eruption starts the cycle again.  Hibernation provides a nice and physically-compelling explanation for a wide array of measures.

Hibernation (and other schemes for CV evolution) operates on times from decades to millennia and longer.  So most of the ways to test Hibernation require somehow getting data from many decades before the nova event to look for changes associated with the evolution.  This is difficult because there is not much century-old data are in existence, and because the CV systems were not watched {\it before} the novae called attention to them.

Fortunately, a solution exists.  A wonderful source of very old data exists in the collections of archival sky photographs, covering the entire sky going deep, at a number of observatories.  These photographs are recorded on emulsion attached to glass plates, and these plates are of the same quality as when they were first recorded.  Large archival plate collections are now at observatories of Harvard, Sonneberg, Asiago, and the Vatican.  The largest of these collections is at Harvard College Observatory, featuring roughly half a million plates from 1889--1954 and around 1969--1989, covering the entire sky, north and south, with any star of 14th mag recorded on thousands of plates, any star of 16th mag recorded on hundreds of plates, and many stars of 18th mag recorded on dozens of plates.  On each of these plates, the brightness of the target star can be easily and accurately measured by comparison with nearby stars of known magnitude.  With this, it is easy to get a century-long light curve for all but the faintest CVs.

A very long light curve can be used to test Hibernation in two ways.  First, we can measure the orbital period of some nova {\it before} the eruption, $P_{pre}$, along with various ways to measure the period {\it after} the eruption, $P_{post}$, to see whether $\Delta$P is really positive.  If $\Delta$P$<$0, that is, if the orbital period decreases across the nova event, then the Hibernation mechanism is not working.  Second, we can test the Hibernation prediction that post-eruption light curve should be fading fast due to the sudden disconnection of the binary turning off the accretion.  Further, century-long light curves can be used to test predictions of other CV evolution models.

This whole enterprise was started back in 1983 (Schaefer \& Patterson 1983), when two of us used the Harvard plates to discover $P_{pre}$ for the classical nova BT Mon (Nova Mon 1939).  We found that $\Delta P/P$ showed an increase in period by 40 parts-per-million (ppm), and we derived that the mass of the material ejected by the nova was 3$\times$10$^{-5}$ M$_{\odot}$.  This would represent the first ever reliable measure of the ejecta mass from any nova, being based on a dynamical measure rooted in well-known physics and accurate/reliable eclipse timings.  (This might surprise some CV theorists, because they have been taking the many published ejecta masses at face value, whereas the real uncertainties in all the published values, both observational and theoretical, are more than two orders of magnitude.  See Appendix A of Schaefer 2011 for a full account of the huge systematic problems that had generally been ignored in publications.)  But the most important implication was that this measure showed that the BT Mon binary system separated significantly during the nova event, so the accretion rate long after the event is completed must also drop.  This result was a substantial part of the original motivation and mechanism for the Hibernation model.

Until recently, this one measure of $\Delta$P for BT Mon has been all that is known.  The problem is that few novae have sufficiently-large photometric effects tied to the orbital period (e.g., from eclipses or ellipsoidal modulations), with a well-placed nova date and a bright enough quiescent magnitude to allow for detection on many plates.  Then, in 2017, we were able to pull out accurate and confident measures of both $P_{pre}$ and $P_{post}$ for the bright V1017 Sgr (Nova Sgr 1919), with its longest-known period for any CN (Salazar et al. 2017).  We reported $\Delta P/P$ of -273$\pm$61 ppm, which is to say that the period {\it decreased} across the nova event.  This is startling both because $\Delta$P is negative and because there is no ready explanation for how the period change can be so large (in any direction).

Over the past several years, we have been pushing deeply and widely for more measures of $\Delta$P.  In this paper, we report on our measures of the light curve of QZ Aur (Nova Aurigae 1964) from 1912--2013.  We use 102 magnitudes from archival plates (including 60 pre-eruption plates) from Harvard, Sonneberg, Asiago, the Vatican, and Palomar, plus 1215 magnitudes from 2009--2016 with CCD measures from several observatories.  Our original goal was to measure $\Delta$P to determine the mass of the ejecta.  Instead, what we found was surprising to us, with our results providing a variety of deep implications for models of CVs and their evolution.

\section{Observations}

QZ Aur was discovered to be a nova in 1975 when N. Sanduleak examined an objective prism plate taken in November 1964 to find an emission line source (Sanduleak 1975).  Gessner (1975) examined Sonneberg archival plates, with the target being very faint before 19 January 1964, brightest at 6.0 mag on 14 February 1964, and fading fast until it was at 15.0 mag on 2 October 1964, with a light curve class of S(25).  Szkody \& Ingram (1994) took spectra on two nights in 1992 with the Kitt Peak 4-m telescope, estimated $P$=0.3575 days, and with their radial velocity curve estimated the white dwarf mass from 0.93--1.02 M$_{\odot}$ and a mass ratio of 0.90--1.00.  Campbell \& Shafter (1995) discovered deep eclipses in their BVRI CCD images with the Mount Laguna 1-m telescope, pulling out an accurate period of 0.3574961 days.  Shi \& Qian (2014) reported eight more eclipse times from 2008--2013.  This short summary represents all that has been published on QZ Aur.

The eclipses are 1.45 mag in amplitude in the B-band, which should be easy to detect on archival plates for a system that has a period of 0.357 days.  The trouble is that QZ Aur has a quiescent magnitude of B=17.50 mag (Campbell \& Shafter (1995), and this faintness means that few plates worldwide can show the system, and ones that do will be near the plate limit.  The critical dataset for our science is the pre-eruption archival plates.  For this, we have made two trips to Harvard and two trips to Sonneberg, plus one visit to Asiago and to the Vatican plate collection now located just outside Castel Gandolfo.  The first Palomar Observatory Sky Survey also has a pre-eruption magnitude from late 1954, while the second sky survey has a post-eruption magnitude in late 1986.  (We also visited the plate collections of the Bamburg, Maria Mitchell, and Jena observatories looking for deep plates showing QZ Aur, but no useful images were found.)  From this, we find 4 pre-eruption and 5 post-eruption eclipses of QZ Aur. 

Full details and procedures for the extraction of modern B-magnitudes are given in general in Schaefer (2016a; 2016b) and Schaefer \& Clayton (2018), while full details and explanations for the case of pre-eruption nova light curves are given in Schaefer \& Patterson (1983) and Schaefer (2011).  The basic data extraction method is the long-traditional direct by-eye comparison of the size of the target star image versus the sizes of many nearby comparison star images of known magnitude.  For the case in hand, with QZ Aur usually being close to the plate limit, this by-eye method is substantially more-accurate and more-reliable than any digital scanning and automated photometry routine.  The light curve is forced to be in the B-band because the plates are all very close to the color sensitivity of the Johnson B-band (indeed, the Harvard plates provided the definition of the B-band) and we used modern B-band magnitudes of the comparison stars.  The result is that we can derive fully-modern B-band light curves with good and useable accuracy.

The error bars on these archival magnitudes are difficult to determine individually.  With our vast experience, we would estimate a typical one-sigma uncertainty of 0.20 mag, which is larger than usual for good plates and good sequences due to QZ Aur often being near the plate limit.  With three independent measures for each Harvard and Sonneberg plate, the average RMS scatter is 0.11 mag, which is a measure of the uncertainty from the measurement error.  The best measure of the average uncertainty is from the RMS scatter of the magnitudes far from any eclipse (taken to have an orbital phase between 0.10 and 0.90) as 0.18 mag.  We adopt this as the real one-sigma error bar.  

This accuracy is not as good as `CCD accuracy' of $\sim$0.01 mag, but it is still more than adequate for the science question at hand.  That is, the photometric accuracy is greatly smaller than the eclipse amplitude, so we can confidently recognize when a plate is or is not inside an eclipse.  Further, this measurement error is comparable in size to the ordinary variations of the flickering, as seen by Campbell \& Shafter (1995).  With the ubiquitous flickering and cycle-to-cycle shape-changes in the eclipse profile, CCD accuracywould not provide any more useful information than what we already have.

A pair of Vatican plates (numbered A1765B and A1766B) both show QZ Aur with nearly-equal magnitudes at the brightness level of the deepest part of the eclipse.  These are a pair of identically centered plates taken with the double astrograph, a pair of co-aligned telescopes on the same mount designed for taking simultaneous exposures.  The log books and other records have now been lost, but the plate jackets give these two plates as being 45-minute exposures, with the times taken back-to-back with a separation of 49 minutes in their mid-exposure times.  But there must certainly be an error somewhere, because two plates separated by 49 minutes cannot both record QZ Aur at its deepest eclipse.  Such errors (in either timing or magnitude) happen occasionally with archival light curves (with a similar rate as for modern CCD observations), with examples including the in-eclipse magnitude at phase 0.19 in the BT Mon light curve (Schaefer \& Patterson 1983) and even the too-bright magnitude (B=16.34) from a Sonneberg plate for QZ Aur.  There are two possible explanations, either of which is reasonable, easy, and possible:  First, as the two plates both are for the same in-eclipse magnitude, they are actually a pair of simultaneous plates, with the time of one of the plates mis-labelled on the plate jacket.  Such minor mishaps in the darkroom happen on occasion, with a moderate frequency.  This is a natural explanation given that the telescope is a {\it double} astrograph, designed for taking pairs of simultaneous plates with identical centers.  Second, the second of the Vatican pair has a slight imperfection of focus plus a 2:1 elongation in a particular direction, such that the image of QZ Aur partly overlaps with the only nearby star.  On other plates, the companion star is well resolved and the images are well separated.  But this combination of circumstances for the second plate makes it possible that the two images interact and overlap so that the original magnitude is wrong.  For whatever reason, one of the two plates of the Vatican pair has either a wrong magnitude or time.  All we can do is report the times and magnitudes as given.  In practice, our analysis below will come to the same result whether we toss out or keep either or both of these two Vatican plates.

We also observed CCD time series photometry with three telescopes, all near Tucson Arizona.  The first run was with the MDM 2.4-m telescope on Kitt Peak, on the night of 15 November 2009, getting a series of 206 images with 10-second exposures in the V-band, all covering one entire eclipse of QZ Aur.  The second was with the Kitt Peak National Observatory (KPNO) 0.9-m, on the nights of 9 and 11 December 2011, getting triples of 180-second exposures in the B-, V-, and r'-bands.  The first of these nights fortuitously covered part of an eclipse.  The third run was with the Steward 61-inch telescope on Mount Bigelow, on the nights of 9 and 10 November 2013, getting time series with 154 images in total that covered two eclipses in the B-band.  The image acquisition, processing, and analysis were all done with standard and well-known procedures (e.g., Howell 2000). 

We also observed QZ Aur with CCD imaging on relatively small telescopes.  These observations are archived in the {\it American Association of Variable Star Observers} (AAVSO) database, with `BDG' for David Boyd, `ARJ' for James Arnold, and `RMU' for Miguel Rodr\'{i}guez Marco.  The last two return a total of 5 V-band magnitudes away from any eclipse.  The Boyd observations are  long time series from 16 nights over 6 observing seasons from 2011/2012 to 2016/2017, with all 831 images taken by an unfiltered CCD where the comparison star V-band magnitudes are used for the differential photometry.  This unfiltered condition means that the resulting magnitudes will have some unknown offset from the Johnson magnitude system, so these cannot be used for following the long term evolution of the system at maximum light.  (Further, these time series were tightly confined in time to the eclipse intervals, so few out-of-eclipse measures were made.)  However, this photometry is good for getting exact eclipse times from 2012--2016.

The comparison stars for all our measures had their standard magnitudes taken from the lists of nearby comparison stars provided by the AAVSO through their {\it AAVSO Photometric All-Sky Survey} (APASS) program.  These magnitudes are based on the usual Landolt standard stars (Landolt 1992; 2009; 2013), thus putting our magnitudes into the Johnson B and V magnitude systems.  For example, the primary comparison star for the MDM magnitudes is an APASS star at J2000 05:28:40.871	+33:19:03.75, with V=13.75 and B-V=0.51.  For the photometry on archival plates, we needed a sequence of comparison stars spanning the entire range of the eclipse, and these B-band magnitudes were taken from APASS, plus an extension privately supplied to us by A. Henden in 2008.  The most important comparison stars for the archival magnitudes (i.e., the very close stars with a magnitude closely equal to QZ Aur at maximum light) are three stars with B-magnitude from 17.05--17.16, all within 30 arc-seconds from QZ Aur.  The three stars have B-V colors of 0.75--0.89, while the brighter primary comparison star has a color index of 0.52 mag.  The color of QZ Aur outside of eclipse is B-V=0.52, and at eclipse minimum B-V=0.67.  Thus, the color difference between QZ Aur outside of eclipse and the primary comparisons is 0.00 mag for the CCD observations and 0.23--0.37 mag for the archival plate sequence.  So we know that any color terms in the systems must be negligibly small.

In Section 7, we will look at whether there could be some offset between the modestly bright comparison stars used for the CCD differential photometry and the faint stars used for the comparison sequence for the archival plate measures.  To test this, we have taken our B-band CCD images with the KPNO 0.9-m and performed ordinary differential photometry on all the comparison stars.  If we take the MDM primary comparison star to have its APASS magnitude, then our differential photometry gives B-magnitudes for all the comparison stars.  We find that the faint comparison stars are accurately given by the old sequences that we used for measuring the archival plates.  In particular, for the three sequence stars near the at-maximum magnitude, the errors in the adopted magnitudes are +0.04, -0.04, and +0.03 mags.  This means that our comparison stars are correct, and they suffer no bright-to-faint offset.

The Palomar second sky survey plates have a significant color term to the B-magnitude system (Reid et al 1991).  So our magnitude comparisons with nearby stars have accounted for the colors of our target star and each comparison star individually, as derived by Johnson et al. (2014).

We have also pulled out two sets of magnitudes from the literature.  Szkody (1994) presents BVR photometry from 2 September 1988 (with the exact time not being stated), plus JK infrared photometry from 30 March 1989, all on the KPNO 1.3-m.  Campbell \& Shafter (1995) presents average magnitudes at maximum and in deepest eclipse for the BVRI filters, (with the exact time ranges going into the averages not being stated), all with the Mount Laguna 1-m telescope.

From all this, we have 1317 magnitudes of our own measure reported here for QZ Aur.  The observations that are not part of long time series are listed in Table 1.  The long time series observations of 2009 (from the MDM observatory) and 2013 (from Steward Observatory) are presented in Table 2, with most of the lines appearing only in the on-line version of this paper.  The long time series from 2012-2016 by BDG is permanently and publicly available on-line in the AAVSO database{\footnote{https://www.aavso.org/data-download}}.  For Tables 1 and 2, the first two columns give the time of mid-exposure, as a heliocentric Julian date (HJD) and as a fractional year.  The next two columns give the band and the measured magnitude in that band.  The final column is the source of the image, where the observatory is named with either the plate number or the telescope aperture quoted in parentheses.

\section{Eclipse Times}

Campbell \& Shafter (1995) report 10 times for eclipse minima, expressed as the HJD of mid-eclipse.  They report that the RMS scatter about their best fit ephemeris is $\sim$100 seconds.  Shi \& Qian (2014) report 8 eclipse times, expressed as the HJD of mid-eclipse, as determined by parabolic fits to the minimum light curves.

We have many well-observed time series covering the entire eclipses.  To get the time of minimum, we fit a parabola to the light curve near minimum.  The formal uncertainties in the times are 8 seconds and 15 seconds for the two KPNO runs, and are 25 seconds to 52 seconds for the BDG time series.  Similar and smaller parabola-fit errors are reported by Shi \& Qian (2014).  However, for measuring the time of the orbital conjunction, we have the additional error caused by the cycle-to-cycle flickering in the source, with this randomly pushing the fitted time of minimum either a bit earlier or later than the time of conjunction.  This flickering jitter is ubiquitous, and there is nothing that anyone can do about it, so this scatter must be added in quadrature to our much-smaller measurement error in the time of the minimum to get the total error in the time of the conjunction.  For this, we take the $\sim$100 second scatter from Campbell \& Shafter as representing their real total uncertainty in the time of conjunction.  We adopt this as the real uncertainty for the two KPNO time series and the Shi \& Qian eclipse times.  Similarly, the RMS scatter for the BDG observations is 130-seconds, which we take to be their total uncertainty.

We have two KPNO time series where the ingress into the eclipse is covered, but the time of minimum is not covered.  We have used our complete light curves to determine offsets from our observations to the time of minimum.  This can only be done with poorer accuracy as compared to a full eclipse profile.  We estimate a one-sigma accuracy of 0.003 days (260 seconds) on these times.

We also have ten plates before and after eruption that certainly show QZ Aur in eclipse (i.e., significantly fainter than the median magnitude, even with the usual flickering), and the time of the mid-exposure is close to the time of the eclipse.  The eclipse plates have exposure times of 45--60 minutes, while the total duration of the eclipse is 72 minutes, so these plates show an average flux over the exposure, which will necessarily be somewhat brighter than the true instantaneous minimum.  For some of the plates, they are close to the expected magnitude at minimum, so we take the plate's mid-exposure time as the best estimate of the time of the eclipse minimum, with an appropriate uncertainty.  For the pair of Vatican plates, we will list the eclipse time as simply being the midpoint.  For the first two of the post-eruption eclipse plates, the magnitude is far from the minimum, so there must be some offset between the observed time of mid-exposure and the needed time of minimum light.  For the post-eruption times, the ambiguity between the ingress and egress branches are easily resolved.  With this, we have used our modern B-band light curve to estimate the time offset and its uncertainty.  For the pre-eruption eclipse-plate times, we have put in no offset because we do not want to make any presumptions as to the shape of the O-C curve.

Table 3 presents all our observed eclipse times.  The first column gives a label in the form $T_i$, with the integer subscript identifying the eclipse times in sequential order.  The next two columns give the times of minimum, first as a heliocentric Julian date (with one-sigma error bars), and second as a fractional year.  The last column gives the source for our times.

\section{Post-Eruption Orbital Period}

Our $P_{post}$ value is derived from the eclipse times in Table 3.  We constructed the usual $O-C$ curve as based on the ephemeris given by Campbell \& Shafter (1995);
\begin{equation}
    T_{linear} = 2448555.1595 + 0.3574961 N_{linear}.
\end{equation}  
The $O-C$ is just $T_i - T_{linear}$.  Figure 1 shows the $O-C$ curve for the post-eruption eclipse times.  The $O-C$ curve shows definite curvature, in the sense that the orbital period in quiescence is decreasing.  As time goes on, the period get smaller, and the orbit must get smaller also.  This should drive an increasing accretion rate.

We ran a simple chi-square fit of a parabola to the O-C curve, with the model ephemeris giving the HJD of minimum as 
\begin{equation}
    T_{model} = E_0 + NP_0 + 0.5N^2\dot{P}.
\end{equation}  
Here, the fiducial epoch is $E_0$ (in HJD), the orbital period at that epoch is $P_0$, N is an integer that represents the cycle count from the epoch, and $\dot{P}$ is the steady rate of period change in units of days per cycle.

Our best fit has $P_0$=0.35749621$\pm$0.00000005 days, and an epoch of HJD 2448555.1591$\pm$0.0002.  We find a $\dot{P}$ value of $(-2.5 \pm 0.5)\times 10^{-11}$ days per cycle.  The negative sign shows the period to be decreasing.  For this fit, we have selected the epoch close to that used by Campbell \& Shafter, as this choice near the middle of the post-eruption eclipse times minimizes the correlations between the fit values and their errors.  The full set of fit parameters is presented in Table 4.

The existence of the $\dot{P}$ term is readily apparent from Figure 1.  Quantitatively, the $\chi^2$ is 69.2 for the best fitting line with $\dot{P}$=0, with a change of 24.0 when adding one fit parameter, so the significance is 4.9-sigma.

With a highly significant $\dot{P}$ term, $P_{post}$ changes, so for calculating $\Delta$P, we have to specify the value for just after the time of the nova eruption.  Gessner (1975) finds that the first plate in outburst was for 14 February 1964, so the relevant time is close to HJD 2438440, with $N_{linear}$ from Equation 1 around -28294.  With this, the $P_{post}$ value is 0.35749691	$\pm$0.00000019 days. The corresponding epoch for the time of the eruption is HJD 2438440.1514$\pm$0.0034.  With the post-eruption ephemeris being known with high accuracy, and with the $O-C$ curve necessarily being continuous across the nova event (i.e., the companion star does not miraculously jump around the orbit), this epoch at the start of the nova event could serve as an eclipse time for us in finding $P_{pre}$.

Our $P_{post}$ measure has been made only with the eclipse data for QZ Aur, but we also must check that this period is consistent with the many observations of QZ Aur near maximum brightness.  Specifically, we must test whether any of the at-maximum magnitudes occur during the orbital phase range of the eclipses.  That is, if we plotted the at-maximum magnitudes as a function of phase, then they should be spread out evenly with no gaps across all phases.  But if we have the correct period, then the phase plot of the at-maximum magnitudes must show a gap for the phase and duration of the eclipse.  A more general test is simply to plot the phased light curve and see whether we get a good eclipsing binary shape similar to the shape known from recent CCD time series.

Figure 2 shows the light curve for all the isolated magnitudes (as in Table 1) phased together.  (The CCD time series from Table 2 reproduces the eclipse shape at the right phase, which crowds the plot, so they are not shown here.)  The Asiago V magnitudes have been converted to B by adding the B-V=0.52 color from Campbell \& Shafter (1995).  Except for six B magnitudes from KPNO in 2011, all 51 magnitudes are from 1967--1990.  This is a relatively short interval, during which the period does not change significantly for the purposes of this plot.  In Figure 2, we see that the in-eclipse magnitudes are clustered tightly around a phase of zero (and its mirror values at phase 1.0 and 2.0), while the at-maximum magnitudes have a blatant gap during the eclipse phases.  This is confirmation that we have the $P_{post}$ value correct.

Our set of eclipse times has a substantial gap from 1992.8 to 2009.8.  This allows for the possibility that the O-C curve in Figure 1 could deviate from the simple parabolic fit over the entire post-eruption fit range.  One possibility is that the O-C curve can have one parabola before 1992 or so, and another after that year, which would imply a singular and large period change for no known reason in the middle of quiescence.  A broken-parabola is the simplest possibility for the exercise of sketching curves to connect 1992.8--2009.8, but in principle we cannot prove any curve within this interval.

As an exercise, we have fitted a broken-parabola to all our 36 post-eruption eclipse times.  This model has six parameters (the pre-break period, the pre-break $\dot{P}$, the year of the break, the epoch at the break, the post-break period, and the post-break $\dot{P}$).  We can indeed fit such a broken-parabola in Figure 1, and the chi-square is lower than for the single-parabola fit.  For a break in 1990, we get a $\dot{P}$ value of around +1$\times$10$^{-11}$ days per cycle, which has the orbital period slowly increasing.  But this lowering of the $\chi^2$ is expected for any addition of three new fit parameters, and the addition of the three fit parameters is not required by the improvement in chi-square.  We prefer the single-parabola possibility over the broken-parabola possibility for two reasons.  First, the unbroken parabola is much simpler than the broken-parabola, and so it is to be preferred by Occam's Razor.  Second, we understand the astrophysics of a constant $\dot{P}$ during quiescence, whereas a sudden period change decades after the eruption has no manifestation in the light curve and no expectation or physical understanding for such a change.  

Another reasonable departure from a simple parabola would be a superposed sinewave.  Roughly sinusoidal deviations in the $O-C$ curve could arise from the effects of activity cycles on the secondary (Applegate \& Patterson 1987), or third-body effects of a star in orbit around the inner binary.  To search for this, we have made a Fourier transform of the residuals from the best fit parabola.  With this, we see only some artifact peaks at very long periods and near one-year, plus lower noise peaks.  The maximum amplitude for a sinewave is 0.002 days, which is representative of the scatter in the residuals.  So we can place a limit of 0.002 days for any sinusoidal term.

The existence of a broken-parabola or sinewave (or any other curve drawn from 1992.8--2009.8) does not matter at all for the basic question of measuring $\Delta$P.  The primary reason is that the $P_{post}$ value for 1964 and the $E_0$ value for the eruption are determined by the 1973--1992 eclipse times, and these do not depend on whatever happens after 1992.  That is, any different conditions after some hypothetical break around 1992 has little to say about the time before the break, while the pre-1992 eclipse times are best for estimating $P_{post}$ in 1964, and that remains close to the values given in Table 4.  A secondary reason is that the exact value of the post-eruption $\dot{P}$, which we will adopt as the pre-eruption $\dot{P}$, is of negligible importance for fitting the pre-eruption light curve.  That is, $P_{pre}$ and $\Delta$P are not sensitive to which value for $\dot{P}$ is adopted.

In all, we adopt the single-parabola fit as shown in Table 4.

\section{Pre-Eruption Orbital Period}

The critical task of our work is to measure $P_{pre}$.  We only have 60 pre-eruption magnitudes and four observed eclipses.  So we have to consider possible issues such as aliasing and the significance of the periodicity.  We will proceed first by looking only at the pre-eruption data, as these will be independent of any post-eruption data.  For this, we derive $P_{pre}$ with three separate methods, with the aim of showing that the result is robust.  Then, in Section 6, we will get the best and final measure of $\Delta$P from a joint fit to all the pre-eruption and post-eruption data.

The pre-eruption period must be fairly close to $P_{post}$, as too large a change in period is not possible.  We take the physically possible range to be within 1000 ppm of $P_{post}$.  That is, we are only looking for periodicities of 0.35713$<P_{pre}<$0.35786 days.  (Actually, we are search over a much broader range yet, see Section 5.2.)  We find that the value of $\dot{P}$ does not affect the pre-eruption eclipse times significantly (because their measurement errors are relatively large and total range of years is relatively small), so we can assume either the post-eruption $\dot{P}$ or zero within no meaningful difference.

\subsection{From the Eclipse times}

We can use the eclipse times $T_{-4}$, $T_{-3}$, $T_{-2}$, and $T_{-1}$ to find all possible orbital periods.  For a linear ephemeris, we can derive the orbital period as
\begin{equation}
    P_{pre} = (T_i-T_j)/(N_i-N_j) = \Delta T_{ij}/\Delta N_{ij}.
\end{equation}  
The idea is to use the eclipses spaced closest in time to give an acceptable range of periods, then to use larger $\Delta T_{ij}$ values to find acceptable periods for any cycle count between the eclipses ($\Delta N_{ij}$).  By working from the smallest to largest time intervals between eclipses, we can narrow down the possible values of the period.

The shortest time interval is from $T_{-2}$ to $T_{-1}$ from plates on two successive nights, which equals 1.0899$\pm$0.0219 days.  There is no ambiguity in the cycle count, so $N_{-2,-1}$=3.  With this, $P_{pre}$=0.3633$\pm$0.0073 days.  This merely confirms the crude range we already have in the previous section.

The next shortest time interval is $\Delta T_{-4,-3}$, equal to 708.0388$\pm$0.0197 days.  The only possible values of $\Delta N_{-4,-3}$ are 1979, 1980, 1981, and 1982.  Then, the only possible periods are 0.357776$\pm$0.000010, 0.357595$\pm$0.000010, 0.357415$\pm$0.000010, and 0.357234$\pm$0.000010 days.  Note that there are no significant difference if we had used either of the Vatican plate times instead of their midpoint.

Importantly, the post-eruption period is certainly rejected, and even any relatively small $\Delta$P is completely rejected.  From above, the appropriate $P_{post}$ can be either 0.3574961, 0.35749636, or 0.35749758 days.  Then, $\Delta T_{-4,-3}/P_{post}$ varies from 1980.536 to 1980.544, which is to say that the observed time interval is greatly inconsistent with any plausible $P_{post}$, which shows that the orbital period changed greatly across the nova event.  To the 3-sigma level, we can reject the possibility that -188$< \Delta P /P <$+150 ppm.  So we already know that the QZ Aur period must have changed by a huge amount.

Similarly, we can use $\Delta T_{-3,-1}$ values to find possible periods.  Over our range, $\Delta N_{-3,-1}$ can be any integer from 10074 to 10094.  But most of these possibilities are rejected by not being compatible with the period possible from $\Delta N_{-4,-3}$.  Only one period is possible for each of the four periods from the previous paragraph.  So for example, $\Delta N_{-3,-1}$=10081 yields a period of 0.3576018$\pm$0.0000021.  In this case, cycle count differences of 10080 and 10082 are outside the acceptable 3-sigma range.  Only in the case of 10091 can a second possibility be fit into the 3-sigma range.  So the only possible values of $\Delta N_{-3,-1}$ are 10076, 10081, 10086, 10091, and 10092.

Our longest inter-eclipse interval is $\Delta T_{-4,-1}$ at 4313.0207$\pm$0.0170 days, for possible cycle counts from 12053 to 12076.  For $\Delta N_{-4,-1}$=12061, the period is 0.3576006$\pm$0.0000014, with this being compatible with all the other requirements from other intervals.  We are left with five possible values of $\Delta N_{-4,-1}$, 12055, 12061,12067, 12073, and 12074.  The last two possibilities are not to be preferred, as they require 2-sigma deviations.  

So we are left with five possible periods, each specified with a very small error bar.  From the eclipse times alone, we cannot determine which of these five possible periods is correct.  However, we can uniquely break the ambiguity by looking at the phase distribution of the magnitudes at maximum light.  That is, at the correct period, the maximum magnitudes will not be inside a phase interval with the eclipse times, while at the wrong periods the maximum magnitudes should fill in the phase interval with the eclipse times.  This is a clear test for the correct period.  For $\Delta N_{-4,-1}$ values of 12055, 12067, 12073, and 12074, within their allowed 3-sigma rages, the phase intervals with the eclipse times are all filled with many maximum magnitudes.  Thus, four of the five possibilities are proven wrong.  Similarly, the $\Delta N_{-3,-1}$ values have all but one of the possibilities rejected by the at-maximum values filling the eclipse duration.  The  $\Delta N_{-4,-1}$=12061 case proves to have a complete gap of all maximum magnitudes over a range corresponding to the eclipse duration (see Figure 4).  That is, only the 12061 cycle count produces consistent eclipse times and a gap in the maximum magnitude during the eclipse.

We come to the identical conclusion (with a slightly larger uncertainty) for $\Delta T_{-3,-1}$.  So this demonstrates that the Vatican plate pair is not needed for a robust period.

So we have solved the $P_{pre}$ question.  The correct period has $\Delta N_{-4,-1}$=12061 and $P_{pre}$=0.3576006$\pm$0.0000014.  This period is robust for any plausible change or problem, and this exhaustively scans a very wide range of positive and negative $\Delta$P values.  So we have our answer.  And the $\Delta P$ value is -0.0001030$\pm$0.0000014 days.  The period has {\it decreased}, with $\Delta$P/P close to -288 ppm.

\subsection{From a Periodogram}

A periodogram is a plot of some statistic sensitive to the shape of a folded light curve that is calculated exhaustively over some range of periods, where the best period is revealed by the periodogram function reaching an extremum.  Schematically, the statistic is calculated for a range of frequencies with a spacing equal to the Nyquist frequency, but in practice we perform the calculations for a very oversampled grid of periods.  A very important case of a periodogram is the Fourier transform, where the Fourier power is calculated for a set of frequencies spaced apart by the Nyquist frequency, where the correct period is revealed by the peak with maximum power.  (The Fourier transform is optimal for finding periodicities with sinewaves with an unknown phase, but it is far from optimal for identifying periodicities with the shape of an eclipsing binary.)  Using a periodogram for the QZ Aur pre-eruption period search has the big advantage that it is exhaustive at trying all trial periods, and we can easily see the structure of the aliases.

The statistic in the periodogram comes in many various forms.  Examples are statistics that measure the dispersion in the folded light curve and that measure the line-length of the connected dots in a phased light curve.  We have adopted a statistic that is sensitive for picking out periodicities whose folded light curve has all the brightest/faintest magnitudes placed into one bin, just the case for an eclipsing binary.  Specifically, we mean-subtract the magnitudes, fold on the trial period, bin the folded light curve, then calculate the sum of the squares of the binned values.  This statistic ($F$) will be largest for the period where all the in-eclipse magnitudes are in the same bin and all the at-maximum magnitudes avoid that bin.  For a trial period that equals the true period, we will have all the eclipse magnitudes in one bin, with no at-maximum magnitudes, creating one bin that has the maximal size, and this value squared will produce a large value of $F$.  If we have a completely wrong period, then the eclipse magnitudes will be spread around in phase, so the mean-subtracted light curve will have all of its binned magnitudes being small, and $F$ will be small.  If we have an alias of the true period, where the eclipse magnitudes all line up, but where the at-maximum magnitudes cover the phase bin with the eclipses, then the `eclipse bin' will be much smaller than the eclipse amplitude and the $F$ value will be large, yet greatly smaller than for the case of the correct period.  

This periodogram has three disadvantages:  First, there is no ready method to calculate the uncertainties on the period, nor any ready way to determine the $E_0$ value.  Second, we have no ready method to quantitatively determine the significance of the periodicity.  Third, the periodogram does not require that the eclipse phase correspond to the $E_0$ from the post-eruption $O-C$ curve.  This means that the noise will be relatively high as some beats and aliases will look good until we realize that such a period requires a discontinuity in the $O-C$ curve around the time of the nova event.  The correct $P_{pre}$ will have a large $F$, but other aliases can have their $F$ values larger than deserved due to the eclipse phase greatly disagreeing with $E_0$.  Still, our periodogram has the shining virtues that it is exhaustive (so we know that we are not missing the true period) and we can see the alias structure.

For the periodogram, we have taken both Vatican plates as in Table 1, even though we think that one of them has a small timing error.  With any such error, the true peak will be substantially lowered, and the noise level will be increased.  But no such error can create any substantial peak.  So even with a timing error for one of the Vatican plates, we know that any high and isolated peak must be the true period.

In practice, for QZ Aur, we have 60 pre-eruption magnitudes from 1912--1964.  The limits (e.g., $B>$17.3 for the earliest point) is taken to be an equality (i.e., $B=$17.3 for the earliest point).  The V-magnitudes are converted to B-magnitudes with a color of $B-V$=0.52 for times outside of eclipse (Campbell \& Shafter 1995).  In our phasing, we have not varied the $\dot{P}$, because these effects are negligible given the relatively large uncertainty in eclipse times (from Table 3, the pre-eruption eclipse times are over one order-of-magnitude worse in accuracy as compared to the post-eruption eclipse times) and the shorter interval duration for the eclipse times (1952--1964 for the pre-eruption times versus 1973--2013 for the post-eruption times).  For the particular run shown in Figure 3, we have used $\dot{P}$=-3.9$\times$10$^{-11}$ days per cycle, but this really does not matter as we get essentially the same periodogram as when we set $\dot{P}$=0.  For calculating F, we used three sets of bins, each out of phase by a third of a bin, so that peaks would not be lost due to the vagaries of a bin splitting the eclipse duration.  We oversampled by a factor of 100$\times$, being gross overkill, but we are assured of catching the position and shape of each peak.  We have calculated $F$ for trial periods ranging from 0.357 to 0.359 days, a range far greater then the 1000 ppm maximum shift that we would consider as physically plausible, to confirm that we are not missing any peak outside the 1000 ppm range.

Our periodogram is displayed in Figure 3.  We see one isolated and prominent peak, reaching up to $F$=60.6.  The second highest peak is at $F$=29.9 at 0.357451591 days.  We also see an alias structure, where the eclipse plates are beating against each other, as peaks in $F$ from 10--29.9 that are approximately regularly spaced.  With this, the true $P_{pre}$ is certainly represented by the highest peak.  This peak has a period of 0.35760076 days.  This matches the period from the eclipse times (see Section 5.1).  With this, the orbital period {\it decreased} across the 1964 nova eruption.  For this periodogram method, we find $\Delta$P to be -0.00010318 days, and $\Delta$P/P to be -289 ppm.

It is impossible for any random noise, timing error, or aliasing to produce such a high peak, greatly above the alias structure.  So we have our answer.  $P_{pre}$=0.35760076 days and the orbital period of QZ Aur decreased, by close to -289 ppm.

\subsection{From a Chi-Square Fit to an Eclipse Shape}

The perfect way to solve all three problems with the periodogram is to use a chi-square fit, where all pre-eruption magnitudes are directly compared to a model light curve for an eclipsing binary.  For each magnitude, for each set of $E_0$, $P_0$, and $\dot{P}$, we calculate a phase from Equation 2, then use a model eclipsing light curve to get a model magnitude, and compare the observed and model magnitudes using the usual $\chi ^2$ equation.  The result is that we can get the total $\chi^2$ as a function of the three fit parameters.  The best fit values for all three parameters are their values for when the $\chi^2$ is minimum.  The 1-sigma error bars are the range over which the $\chi^2$ value is within 1.0 of the global minimum.  (This solves the first disadvantage of the periodogram.)  And the significance of the periodicity can be determined from the usual probability tables for the minimum $\chi^2$ value.  (This solves the second disadvantage of the periodogram.)  The phasing of the times according to Equation 2, along with requiring that $E_0$ be consistent with the post-eruption value ensures that the $O-C$ curve is sensibly continuous.  (This solves the third disadvantage of the periodogram.)  This chi-square fitting method allows for the correct use of the limits (e.g., $B>$17.3), it makes no approximation (e.g. by putting the time simply into a bin), and it treats each point individually (without averaging anything).

The model eclipse light curve that we used has an at-maximum brightness of $B$=17.16 lasting from phase 0.05 to 0.95.  At mid-eclipse, the model gets down to B=18.35.  (The archival plates have exposure times of typically 60-minutes, so the observed depth will be somewhat smaller than observed with much shorter integration times.)  The shape of the eclipse profile is piecewise linear so as to approximate our eclipse profiles.  We adopted a symmetric eclipse profile, even though the modern CCD photometry shows some small asymmetries.  Indeed, the CCD photometry shows substantial cycle-to-cycle variations in the eclipse profile (likely cause by the usual flickering suffered by all CVs), so any profile can only be approximate.

The pre-eruption light curve does not place any useful constraint on the $\dot{P}$ value, so we have set  the pre-eruption $\dot{P}$ equal to the post-eruption $\dot{P}$ value of $-3.3\times 10^{-11}$ days per cycle.  Again, this choice is not substantially different from the case if we had a zero $\dot{P}$.  With this, we have just two free parameters; $P_{pre}$ and $E_0$.  We take the $E_0$ value to vary near the epoch at the time of the nova eruption.

We have used our chi-square fits to explore possible values for $P_{pre}$.  We find that the $P_{pre}$ values near 0.35760076 days (see previous section) give by far the best fitted $\chi^2$.  Further, the many other trial periods all have greatly worse $\chi^2$ values.  All this is to say that our third period search technique (the chi-square fit) also uniquely pulls out the $P_{pre}$=0.35760076 days peak as being best. 

Our chi-square fits find a single minimum, blatantly better than all aliases.  Full details are listed in Table 4.  We have $P_{pre}$=0.35760093$\pm$0.00000004 days, for an epoch in 1964 just before the nova event.  Within the uncertainties, this is identical to the value derived from the other two techniques.  The 60 pre-eruption magnitudes are folded on this best fit as shown in Figure 4.  Pointedly, we see a good eclipsing binary light curve with the duration and depth as for the post-eruption light curve, and we see the tight clustering of the in-eclipse magnitudes to a phase range over which there is an otherwise-surprising gap of at-maximum magnitudes.  That is to say, the folded pre-eruption data look good and provide strong confidence in the reality and accuracy of the period.

A test of the existence of this period is whether the epoch $E_0$ from the pre-eruption magnitudes approximately equals the value extrapolated from the post-eruption eclipse times.  That is, the $O-C$ curve must be continuous across the eruption, because the companion star cannot jump around in its orbit.  (The {\it slope} of the $O-C$ curve can change discontinuously, and indeed, that is what we are trying to measure with $\Delta$P.)  The $E_0$ from the pre-eruption magnitudes is completely independent of the post-eruption eclipse times.  Yet the difference in the two values is 0.0022$\pm$0.0034 days.  That is, the epochs from the completely independent pre-nova and post-nova data are the same.  Importantly, uncertainty of this difference is just under 1\% of the period.  There is only a chance of 1-in-104 that these $E_0$ values would agree so well if the $P_{pre}$ value were some sort of an artifact.  So the agreement of the independent $E_0$ values provides substantial evidence for the reality of the pre-eruption periodicity.

The best fit $\chi^2$ is 131.1 for 58 degrees of freedom.  Just two points (the second of the Vatican plate pair and the B=16.34 point) contribute 84.8 to this total $\chi^2$.  Without these two magnitudes, the total chi-square is 46.3 for 56 degrees of freedom.  Both of these magnitudes are always on the flat part of the eclipsing binary model (for relevant trial periods of interest), so changes in the trial period yield no change in the $\chi^2$.  That is, the existence of a timing error for the Vatican plate still leaves us with a significant periodicity, while the rejection of the point will substantially improve the calculated significance.  

The best fit $\chi^2$ of 131.1 can compared to the second best for the highest alias (0.357451591 days) of 164.3.  This is a difference of 33.2, corresponding to the preference for the best fit value at the 5.8-sigma level.  The best fit $\chi^2$ can also be compared to the typical value for non-alias trial periods of $>$350 and with a formal average of 461.  This difference corresponds to a preference for the best fit value at the $>$15-sigma level.  So the existence of the periodicity, and the rejection of the aliases, is highly significant.  (But we already knew this by a brief examination of Figure 3.)  

Our primary science result is based on $P_{pre}$ being greatly different from $P_{post}$, and $P_{pre}$ appears to be based on just five eclipse plates, of which one plate must have some problem.  So how robust is this result?  Well, we have given three analyses by three greatly different methods, each exhaustive, all reaching the same result.  We can also check for the robustness of various assumptions and data selections.  The usual formal test is looking at the improvement in chi-square ($[\Delta\chi^2]$) as one parameter is added to the fit, this being an F-test.  The F statistic is the ratio of $[\Delta\chi^2]$ divided by the reduced chi-square, with the resultant probability showing whether the addition of the extra fit parameter is justified by the reduction in chi-square (Chapter X, Bevington \& Robinson 2003, pages 207-208).  We have three applications to test the robustness of our measured $P_{pre}$.

Our first application of the F-test is to see whether it matters that we assumed $\dot{P}$ to equal the post-eruption value.  So we can compare the case where we have a simple linear fit for the ephemeris of the pre-eruption data versus the case where we add a $\dot{P}$ term.  The chi-square for the two cases is 131.7 (for the linear fit) versus 131.1 (for the parabolic fit), giving $[\Delta\chi^2]$=0.6, an F value of 0.25, and a probability of 0.62 that the data requires the addition of a $\dot{P}$ term.  The moderate probability shows that we are not required to add the $\dot{P}$ term.  More to the point, the small $[\Delta\chi^2]$ value even for a large change in $\dot{P}$ shows that the exact value of $\dot{P}$ does not matter in our analyses for $P_{pre}$.  Again, the reason why the $\dot{P}$ does not matter for the pre-nova light curve (but does matter for the post-nova eclipse times) is that the eclipse timing accuracy is an order-of-magnitude worse while the interval with eclipses is 4$\times$ worse.  So our analyses are robust for any plausible changes of $\dot{P}$.

Our second application is to test for the existence of a periodicity.  We can do an F-test comparing a model with the known eclipse shape and amplitude (resulting in a chi-square of 131.1) versus a flat model light curve (resulting in a chi-square of 265.2).  The cases differ by only one parameter, the amplitude of the eclipse.  (Actually, we just adopted the modern amplitude for our model light curve, whereas we could have varied the amplitude to get a somewhat better chi-square.)  The $[\Delta\chi^2]$ value is 134.1, the F value is 58.2, and the probability of the full-amplitude being not needed is 3$\times$10$^{-10}$.  That is, the existence of our periodicity is highly significant and robust.

We can extend this application by testing whether our result is robust to the deletion of any of the eclipse plates.  This is relevant because one of the two $T_{-4}$ plates must have some sort of a timing error.  So we have applied the same test as in the previous paragraph, except that we have not included the two Vatican plates for $T_{-4}$.  With the two Vatican plates ignored, the chi-square improves by 69.1 from the zero-amplitude case to the full-amplitude case, with F equal to 63.4 and the probability of 1$\times$10$^{-10}$.  That is, our pre-eruption periodicity is highly significant even if we delete both $T_{-4}$ plates because one of them has some problem.  We can further try to ignore each of the $T_{-3}$, $T_{-2}$, and $T_{-1}$ plates in turn.  The resultant probabilities are $<$9$\times$10$^{-9}$.  So, the existence of our pre-nova periodicity is highly significant and robust, even for deleting any of our eclipse plates.

Our third application of an F-test is to check whether the addition of a $\Delta$P parameter is justified by the improvement in chi-square.  That is, how significant is our non-zero $\Delta$P for our highly-significant periodicity?  The case with $\Delta$P=0 (i.e., extending the post-eruption ephemeris to before the nova) returns a chi-square of 344.8.  With the chi-square varying substantially for changes in the ephemeris, it might be better to take the average chi-square for small $\Delta$P values, with this average being 265.  This small-$\Delta$P case is worse than the best-fit free-$\Delta$P case by $[\Delta\chi^2]$=134.0, for a probability of 3$\times$10$^{-10}$.  That is, our periodicity and the large non-zero $\Delta$P is both highly significant and robust.

So, again, we have a highly-significant and robust $P_{pre}$, with this value being substantially larger than $P_{post}$.  

\section{Joint Fit For Pre-Eruption and Post-Eruption Data}

The previous two sections separately examined the pre- and post-eruption data, with the fits treated entirely independently.  But the two fits are actually joined by the commonality of $E_0$.  That is, the $O-C$ must be continuous, so the $E_0$ for the pre-eruption data must equal the $E_0$ for the post-eruption data.  Forcing this equality allows for some trade-off of $E_0$ for $\Delta$P.  Now, we already know that any such effects will be small, because the $E_0$ values are independently close to each other, and the period change is so large and accurately measured.  Nevertheless, we should run a single joint fit for all our data.  This will provide our best and final measure of $P_{pre}$, $\dot{P}$, and $\Delta$P.

To be specific, we run chi-square analyses for the 60 pre-eruption magnitudes and the 28 post-eruption eclipse times.  Further, the model for the phasing is taken from Equation 2 for the pre-eruption magnitudes, while the post-eruption eclipse times are also phased with Equation 2.  For the two uses of Equation 2, we adopt one value for $\dot{P}$ applicable to both before and after the nova.  Further, the two uses of Equation 2 have the $E_0$ value set equal to each other.  Our model with four free parameters ($P_{post}$, $P_{pre}$, $E_0$, and $\dot{P}$) is trying to fit 60 magnitudes and 36 eclipse times, so our joint fit has 92 degrees of freedom.  The 1-sigma range of uncertainty for the fit parameters (and the calculated quantity of interest $\Delta$P) is found by looking over the volume in parameter space over which the $\chi^2$ value is within 1.0 of the minimum value.

Our joint fit gives $\dot{P}$=(-2.84$\pm$0.22)$\times$10$^{-11}$ days per cycle.  The best fit for the joint data set gives $P_{pre}$=0.35760096$\pm$0.00000005 days.  With our $P_{post}$, we have $\Delta$P=-0.00010393$\pm$0.00000010 days, and $\Delta$P/P=-290.71$\pm$0.28.  So again, we find definitively that the orbital period of QZ Aur {\it decreased} by a large amount across its 1964 nova eruption.

The full $O-C$ curve for all the eclipse times (from Table 3) are shown in Figure 5, along with the joint best fit model.  We see that the both the $\dot{P}$ term and the small scatter in the post-eruption timings are negligibly small for questions of $\Delta$P.  We see that the epoch at eruption, $E_0$, is closely constrained by both the full post-eruption curve extrapolated back to 1964.112, as well as the two eclipse plates from Sonneberg at 1964.05.  The cycle count for the pre-eruption plates are known with high confidence from any of the three analyses in Section 5.  Given this robust cycle count, the sharp kink in the $O-C$ curve shows that the $\Delta$P is highly significant, different from zero, and is accurately measured.  The $O-C$ diagram shows that the break is `concave down', and this means that the orbital period of QZ Aur has {\it decreased} across the 1964 eruption.

We should check the significance of the period-change with this joint chi-square fit, as in Section 5.3.  With the F-test for comparison between our best-fit model and the zero-amplitude model, the existence of the periodicity is confirmed again at very high confidence levels, close to those quoted in Section 5.3.  And we again confirm that the significance remains very high if we delete both $T_{-4}$ plates, or indeed for the deletion of any of the pre-eruption eclipse plates.  

This particular F-test is asking the question as to whether the pre-eruption light curve significantly displays an eclipsing binary shape, which is to say whether our pre-nova light curve has a significant periodicity as given.  But we know that the pre-nova system really must show eclipses with something close to the modern eclipse profile.  So there inevitably exists some eclipse period, and we can ask what is the significance of our $P_{pre}$ versus some other period.  This question can be addressed with an F-test comparing our best-fit $P_{pre}$ versus a model light curve for which only the pre-eruption period is allowed to change.  Another way of stating this question is whether our best-fit $P_{pre}$ can be proven to be inconsistent with being a random shuffling of magnitudes in a phase plot so as to mimic an eclipsing light curve.  The most visible part of this is simply whether the plates with in-eclipse magnitudes all tightly cluster within some range of phase about 0.08 wide.  But an equally important part is that the at-maximum magnitudes must have a gap in the {\it same} phase range.  (See for example in Figure 4, where the phase range around 0.0, or 1.0, has no out-of-eclipse magnitudes, while the four plates in the gap all show QZ Aur deep in eclipse.)  An at-maximum plate inside the gap provides as much chi-square as does an in-eclipse plate outside the gap.  The F-test of the zero-amplitude case does not include this additional information (i.e., that a gap of the at-maximum light curve must appear at the same phase as the in-eclipse points), so we can do better for testing the significance of $P_{pre}$.  We can also do better by using the additional and certain information that the $O-C$ curve must be continuous across the eruption.  This is to say that the eclipse plates (and the gap in the at-maximum plates) must occur at zero phase, not 0.1 or 0.45 or whatever.  This can be viewed as the epoch $E_0$ providing a very-accurate eclipse time ($T_0$ as in Table 3) that the pre-eruption ephemeris must conform with.  We effectively have five eclipse times measured, not four.  So an F-test operating on the joint fit and comparing our $P_{pre}$ with other trial periods will be the best measure of its significance.

Our best-fit $\chi^2$ value is 174.1.  We can ask for the probability that such a low value can be obtained with the best-fit $P_{pre}$ arising from a shuffling of the light curve points at some wrong period.  The $\chi^2$ for periods within 1000 ppm of $P_{post}$ is best made by simply averaging calculated values for the model with the best-fit parameters except that the pre-eruption orbital period is changed within this range.  (This preserves the effects of plates close together in time, such as the Vatican plate pair and the pair of $T_{-2}$ and $T_{-1}$.)  The $\chi^2$ ranges from 256 to 717, with an average of 406.  So the model with our best-fit period improves the chi-square by 213.7 over the model that differs only in having an average nearby period.  With 57 degrees of freedom, the F value is 75.9, for a false alarm probability of 5$\times$10$^{-12}$.  This F-test gives a probability that any one period will produce a chi-square as low as seen for our $P_{pre}$, but we have effectively tested many periods within 1000 ppm of $P_{post}$.  The half-width of the chi-square minimum is around 1 ppm, so we have roughly 1000 tested trial periods from which we selected out the smallest value.  So the F-test probability must be multiplied by 1000 to get the probability that any one of the tested trial periods might have the light curve magnitudes shuffled in time so as to create a false period as good as we see.  So our final probability is 4$\times$10$^{-9}$ that our excellent eclipsing binary light curve can be produced as a false alarm.  That is, it is very very improbable that some false period will shuffle the magnitudes in phase so as to produce a good eclipsing binary light curve with $\chi^2$=174.1.  Thus our $P_{pre}$ is significant at the 4$\times$10$^{-9}$ probability level.

The most important of our observational results is that $\Delta$P/P equals -290.71$\pm$0.28 ppm, and this depends on the four pre-eruption eclipses, for which one of the plates is known to have a error.  So we have to ask whether our $P_{pre}$ is robust on single, double, triple, or even quadruple errors in our pre-eruption plates.  (1) The quick and confident answer is that any error on a single plate (whether in magnitude or timing) could only make the folded light curve worse, so if our light curve has such an error then the true significance of the period can only be better than we calculate.  So, yes, our $P_{pre}$ is robust for singular errors.  (2) Let us also examine the case where both $T_{-4}$ plates are arbitrarily rejected.  Our best fit then produces $\chi^2$=102.8.  The average chi-square is 282 for trial periods within 1000 ppm of $P_{post}$.  The F-test probability is 1$\times$10$^{-13}$.  With 1000 trial periods, our $P_{pre}$ is significant at the 1$\times$10$^{-10}$ probability level.  (3) Let us examine the case where we arbitrarily ignore three plates (for $T_{-4}$ and $T_{-2}$).  This leaves us with just two pre-eruption eclipse plates (for $T_{-3}$ and $T_{-1}$), and the period is lost if we cannot keep the cycle count.  But our analysis in Section 5.1 shows that we can keep the cycle count, because the gap in the at-maximum magnitudes is present for only one possible period.  Further, by using the joint fit, we really have three eclipse times; $T_{-4}$, $T_{-2}$, and $T_0$.  With this, the best fit chi-square is 100.9, the average chi-square for nearby periods is 278, and the final probability of a false alarm is 2$\times$10$^{-10}$.  (4) We can even get an accurate and confident $P_{pre}$ in the case where we arbitrarily reject four of our pre-eruption eclipse plates, leaving only one pre-eruption eclipse.  If we only have $T_{-3}$, then our best fit chi-square is 97.2, while the average chi-square for nearby periods is 239.  This gives a false alarm probability of 6$\times$10$^{-9}$, corresponding to a 5.8-sigma result in a Gaussian distribution.  That is, our $P_{pre}$ is still highly significant if we toss out four-out-of-five of our pre-eruption plates in eclipse.  This result is surprising if you only think that there is one eclipse time, but we also have the $T_0$ plus the requirement that all the at-maximum plates have a gap centered at zero phase.  Effectively what is going on is that the period is $(T_0-T_{-3})/\Delta N_{-3,0}$, with the at-maximum magnitudes eliminating all but the correct value of $\Delta N_{-3,0}$.  This period from just one pre-eruption plate is simple to derive, of good accuracy because $T_0-T_{-3}$ is large, and the gap at zero phase provides a reliable way to uniquely pick out the correct $\Delta N_{-3,0}$.  The point of the analyses in this paragraph is to prove that our $P_{pre}$ is robust even for ignoring many of the eclipse plates.

We have gone into great detail with three methods to derive $P_{pre}$, resulting in a confidence level of 3$\times$10$^{-10}$, plus one definitive method to derive $\Delta$P from the joint fit, and even dropping most of the eclipse plates still yields a highly significant period.  So we have seen a very robust and highly-confident detection of the pre-eruption eclipse period, and we have a highly-accurate measure of that period.  Our $P_{pre}$ measure has no ambiguities, no outstanding problems, and we cannot think of any way to try to impeach our result.  So theorists are confronted with $\Delta$P/P of -290.71$\pm$0.41 ppm for QZ Aur.

\section{Long-Term Light Curve Before and After the Eruption}

QZ Aur is one of a very few CNe to have a long pre-eruption light curve (52 years) plus a long post-eruption light curve (50 years).  This provides a rare opportunity to test predictions by various models of CV evolution.

For this, we have extracted all the magnitudes with the final phase between 0.10 and 0.90, so that we can be confident that there are no eclipse effects.  We have then binned these together based on proximity in time.  The Asiago V-band magnitudes are converted to B with B-V=0.52.  Within each bin, we have averaged the magnitudes together.  The 1-sigma error bar is taken to be the larger of 0.18 mag and the RMS of the included observations, divided by the square-root of the number of magnitudes, all with a lower limit of 0.02 mag.  This lower limit is a small acknowledgement that the irregular flickering in QZ Aur makes for an instance-to-instance variance that contributes to the overall uncertainty of the measure of the average magnitude at maximum.  Our resulting light curve is compiled in Table 5 and displayed in Figure 6.

We see a flat quiescent light curve from 1925 to 1981, excepting 1964.122 to 1970 for the eruption and its tail.  This level is close to B=17.14 mag.  

We have the limit ($B>$17.3) for one Harvard plate from early 1912.  On the face of it, this implies that QZ Aur was below the usual quiescent level back in the 1910s.  But this is not a conclusion with high confidence, as the 1-sigma uncertainty on this limit is the usual 0.18 mag, so QZ Aur might well have been at the usual quiescent level back in 1912.

For both the 1988--1990 and 2009--2014 intervals, we have the conundrum that the light curve has a behavior that is different from the pre-1980s in two ways.  First, the average magnitude is substantially and significantly fainter than the pre-eruption level, being B=17.44 in 1988--1990 and B=17.49 in 2009--2014.  Second, the light curve displays substantial and significant variability, ranging from 17.10--17.98 mag.  So, after 1981, QZ Aur is fading with large fluctuations.  This could be viewed as either QZ Aur fading in some manner below the pre-eruption level starting after 1981, or as QZ Aur being at the pre-eruption level from 1988--2014 yet with frequent dips by up to 0.84 mag or so.  In all cases, after 1981, the {\it average} magnitude for QZ Aur started falling {\it below} the pre-eruption level.

This conclusion (that QZ Aur is fading substantially and well-below its pre-eruption level) has broad implications, so we must question whether this result is robust and confident.  As discussed in Section 2, we are indeed confident in our photometry to within the stated error bars.  But we see that the at-maximum behavior largely changes as our magnitude sources switch over from archival photographs to CCDs.  That is, the archival plates are showing a fairly steady state with B around 17.14, while the CCD measures show $>$2$\times$ variations around an average B of 17.49 mag.  These behavior changes (fairly flat to large variability with an average shift of 0.35 mags) might be associated with the date (pre-1980s versus post-1980s) rather than the measuring instrumentation (plates versus CCD).  If the behavior change comes with the instrumentation, then our conclusion about the post-eruption fading of QZ Aur is not robust.

Realistically, instrumentation problems can arise only from a few well-known possibilities.  One of these possibilities is that perhaps the comparison stars have some systematic offset (at the 0.35 mag level) between the faint stars serving as the sequence for the archival plates and the brighter stars serving as the comparison for differential photometry for the CCD images.  But, all observations are keyed off the single APASS sequence, and our high-signal, good angular resolution images confirm the accuracy of the sequence to within 0.04 mag (see Section 2).  So the problem is not in the comparison stars.  

Another possible instrumentation artifact is the color terms that arise due to each instrument not having exactly the same spectral sensitivity for the Johnson B-magnitude system.  The B-V color differences between the comparison and target stars will lead to a nearly-constant offset between magnitudes measured at different observatories.  The problem would have to come from the CCD observations, because there is no chance that such systematic errors in the archival plates will offset large and random variability to make the archival light curve appear flat.  In principle, these color term offsets can be quite large, and perhaps could account for how the many CCD observers come up with a large scatter.  But in practice, none of this works this way.  Quite varied CCD systems have offsets usually at the 0.05 mag level when dealing with with targets and comparisons with large differences in B-V (Schaefer et al. 2011; 2017).  The color terms for the systems at the professional observatories are designed to be small.  And for the magnitudes reported to be in the B-band by Campbell \& Shafter as well as by Szkody, they are top quality and highly experienced observers, so it is not plausible that they would differ from the B-system by anything like 0.35 mag.  For the case of QZ Aur with archival plates, the target and comparison stars have B-V differences of 0.23--0.37 mag for outside of eclipse, so the color terms will only be a small fraction of the coefficient.  For the case of CCD observations, the target and the primary comparison star have zero color difference, so the color term is zero.  So we see no realistic way at which the CCD observations could possibly have variations in their color terms to explain either the 0.35 mag offset or the $>$2$\times$ variability.  For the archival plates, extensive study and results from the last half century by many workers has shown that the effective spectral sensitivity of the Harvard plates is unchanging and equal to that of the Johnson B-magnitude system.  The plates from the other observatories used the same emulsions and photochemicals, and the experience here is also that the color terms are zero between the plates' native system and the Johnson B-system.  In all, the color terms are certainly negligibly small, and certainly this possible instrumental problem cannot account for variations in the long term light curve of QZ Aur.

Another conceivable instrumental effect might arise from contamination of the QZ Aur light by a B=18.22 star that is 9.7 arc-seconds to the WSW.  Some imaging systems might merge the images, or the photometric aperture might be made so large as to include the light of the companion.  If all the light from the companion star were to be mistakenly added in to the QZ Aur flux, then the expected out of eclipse magnitude would be one-third of a magnitude brighter.  However, all of the CCD images that we are reporting on have image sizes greatly smaller than half the separation, and normal CCD photometry practice would require that the photometry aperture be much smaller than the separation.  For example, our KPNO 0.9-m images have a radius of the full image (down to the background) of 0.8 arc-seconds for QZ Aur, a separation of 9.7 arc-seconds to the companion star, and a photometry aperture of 1.27 arc-seconds in radius, so there is no chance of any contamination.  For the archival plates, they have plate scales similar to the Palomar sky survey images, and we see that QZ Aur is well separated from its companion star.  In particular, the comparisons made between the image sizes of QZ Aur and its comparison sequence stars have no effect from the nearby companion star when the images are not touching.  The only case of relevance is that the APASS images have a relatively large pixel size, so some care must be taken to not blindly have too large a photometry aperture.  But we have not used APASS magnitudes for QZ Aur itself, so this is not an issue for our light curve.  In all, any idea of contamination by the nearby star is certainly wrong, and this instrumental effect cannot account for the fading or variability in our light curve.

So all the known instrumental effects are confidently determined to be greatly too small to cause the behavior change from before the 1980s versus after the 1980s.  There is no evidence for the behavior correlation with instrumentation being in any way instrumental, and we have strong evidence that the instrumental effects cannot be anywhere near large enough to account for our out-of-eclipse light curve.  So we have to take at face value the fading of QZ Aur after the 1980s to significantly below the pre-eruption level.

\section{How Can the Orbital Period Possibly {\it Decrease} By So Much Across an Eruption?}

What could be the cause of the large decrease in orbital period observed in QZ Aur? The orbital period changes in nova outbursts were recently reexamined in a paper by Martin, Livio and Schaefer (2011), henceforth MLS.  They showed that under the conservative assumption that the ejecta carry the specific angular momentum of the white dwarf, the orbital period is expected to increase and that the relative change in orbital period is equal to twice the fraction of mass loss from the system.  Namely 
\begin{equation}
\frac{\Delta P}{P} = 2\frac{\Delta m_1}{M}\ ,
\end{equation}
where $\Delta m_1$, taken to be positive, is the mass ejected from the white dwarf, and $M$ is the total mass of the system.  We shall follow their notation in our subsequent discussion, and follow their treatment casting it in terms of the ratio $r=(\Delta P/P)/(\Delta m_1/M)$. They also consider the effects of mass accretion by the companion, re-visit the frictional angular momentum losses due to the interaction of the binary with the ejecta previously considered by Livio, Govarie, \& Ritter (1991), and add a new mechanism by including the effects of the magnetic field of the secondary star forcing co-rotation within the Alfven radius. They show that these additional effects may reduce $r$ well below the conservative value of 2, and that for low mass ratios $q=M_2/M_1\leq 0.5$, it is possible to obtain $r<0$, in other words, a decrease of orbital period upon mass loss. 

For QZ Aur we may estimate the mass loss due to the nova explosion from the standard critical pressure  $P_{\rm crit}\approx 10^{20}$ dyn/cm$^2$ (Shara 1981; Fujimoto 1982a,b; MacDonald 1983, Starrfield, Iliadis \& Hix 2016) as
\begin{equation}
\Delta m_1 \approx 4\pi R_1^4 \frac{P_{\rm crit}}{G M_1} .
\end{equation}
With $R_1=5.75\time 10^8$ cm, for $M_1=0.98 M_\odot$, this yields $\Delta m_1=5.3\times 10^{-4}$ , and thus $\Delta m_1/M = 2.8\times 10^{-4}$. The observed decrease in orbital period therefore requires $r\approx -1$, which appears virtually impossible for a binary with $q$ close to unity, even with the most generous assumptions about the magnetic field of the secondary (e.g. Fig 2 of MLS).  Since all of the above estimates assume either explicitly or implicitly the ejection to be spherically symmetrical 
around the white dwarf, it makes sense to explore the possible effects of an  asymmetric ejection using a simple model. 

Let us assume that the ejection consists of two hemispherical shells with different masses being ejected forward and backward along a common axis of symmetry which is coincident with  the instantaneous orbital motion of the white dwarf. This assumption is made for simplicity and to maximize the effect of the asymmetry in the ejection without having to account for any radial kick. For spherical half-shells ejected radially with velocity $v_{\rm ej}$ the average projection onto the direction of motion of the white dwarf is $v_{\rm ej}/2$. Let $\Delta m_{\rm f}$ and $\Delta m_{\rm b}$ be the masses of the forward directed and backward directed shells, such that $\Delta m_1 =\Delta m_{\rm f}+\Delta m_{\rm b}$. The net forward momentum carried away by $\Delta m_{\rm f}$ would therefore be $\Delta m_{\rm f}(v_1 + v_{\rm ej}/2)$, whereas the backward shell carries $\Delta m_{\rm b}(v_1 - v_{\rm ej}/2)$, where $v_1$ is the orbital velocity of the white dwarf.  Therefore the net orbital angular momentum change is
\begin{equation}
\Delta J = - a_1\Delta m_1 v_1(1+ \xi) v_{\rm ej}/(2 v_1)\
\end{equation}
and we have set $\xi = (\Delta m_{\rm f} - \Delta m_{\rm b})/\Delta m_1$, a parameter measuring the forward asymmetry, so that $\xi=0$ for the usual assumption of symmetrical nova ejecta and $\xi=1$ if all the mass is ejected in the forward direction. Then, following the steps as outlined in MLS, it is easy to show that the relative change in binary separation  is 
\begin{equation}
\frac{\Delta a}{a} = \frac{\Delta m_1}{M}\left(1- q\xi\frac{v_{\rm ej}}{v_1}\right)\, ,
\end{equation}
and thus the relative change in the orbital period is 
\begin{equation}
\frac{\Delta P}{P} = \frac{\Delta m_1}{M}\left(2- q\xi\frac{3 v_{\rm ej}}{ 2v_1}\right)\, .
\end{equation}
For the parameters of QZ Aur, $v_1 = 188$ km/s, whereas the velocity of the ejecta is not known unequivocally but is probably in the range of 300-3000 km/s. Getting $r = 2- q\xi\frac{3 v_{\rm ej}}{ 2v_1}=-1$ would require an asymmetry $\xi=0.38$ for $v_1=1000$ km/s, and less for a faster nova. While these requirements are non-trivial, they demonstrate that a significant decrease of the orbital period is possible during a nova outburst.

The momentum asymmetry in the ejected shell need not come from a mass imbalance, as it could also arise from a velocity difference with direction.  For a difference in ejected velocity to produce a period-change, it would have to have a monopole component, while a dipole ejection (creating a bipolar shell) will not make for a period-change.  And the period-change could result from combinations of velocity and mass imbalances.  The extreme case would be a jet pointing in the forward direction.

So we can easily explain the large negative $\Delta$P if the nova shell has substantial asymmetry.  Such asymmetries are frequently seen in the ejected nova shells.  The first and best observed nova shell, around GK Per, has 65\% of its knots in a semi-circle towards the SSW, 72\% of its knots towards a hemisphere towards the south coming towards Earth, for $\xi$=0.44, while it also has a narrow single-sided jet (Shara et al. 2012; Liimets et al. 2013).  A sampling of nova shells with well-resolved imaging that turns up in a search of recent papers reveals (1) the shell of V5668 Sgr as seen with ALMA has roughly 80\% of the flux on the southside (Diaz et al. 2018) for $\xi$$\approx$0.6, (2) the nova shell of AT Cnc has roughly 95\% of the flux in a half-circle directed towards the NE (Shara et al. 2017) for $\xi$$\approx$0.9, and (3)  the nova shell around `J210204' has $\sim$95\% of its flux in one hemisphere (Santamaria et al. 2018) for $\xi$$\approx$0.9.  For a more systematic sample, Downes \& Duerbeck (2000) present a survey seeking nova shells, for which half of the detected shells have axial ratios from 1.1--1.42, and half of the shells have apparent mass asymmetries corresponding to non-zero $\xi$ values, up to $\xi$$\approx$1 for V3888 Sgr.  These $\xi$ values are under-estimates as the heavy-mass hemispheres will be pointing to Earth to some extent, but the possibility of a sharp gradient in the ISM density with the edge near the star could make for a change in the $\xi$ value in some rare cases.  Still, it is clear from the nova shells that large values of $\xi$ are common.  And with this, we have an easy and common mechanism to create large $\Delta$P values for many novae.

So we know that large imbalances in the shell ejection provide an easy mechanism to create large period changes, and we have proof that the shells often have large imbalances.  But we do not have a mechanism for creating the imbalances.  We can easily envision that the ejection from the white dwarf could be directional (with respect to the magnetic fields or the companion position).  But such directions will be rotating with the white dwarf or the orbit, so we do not see how the long-ejection can be exclusively in the forward direction of the white dwarf's orbital motion.  But this worry is solved if the ejection imbalance occurs over a time interval substantially shorter than an orbital period.  So we have a reasonable explanation for the large $\Delta$P values.

One implication of the large $\Delta$P values, by any mechanism, is that the period-change is not dominated by the simple loss of mass by the white dwarf during the nova eruption.  Some other effect dominates.  This means that we cannot usefully use something like Equation 4 or 8 to go from an observed period change to a measure of the ejected mass.  (A dynamical measure of the ejecta mass as based on precise timing, and independent of extinctions and distances, is valuable, because all other methods have unresolvable 2--3 orders-of-magnitude uncertainties; see Schaefer 2011.)  This implication is horrifying to us, because we have been working since 1983 on the large-scale program of measuring $\Delta$P for many CNe and recurrent novae (e.g., Schaefer \& Patterson 1983; Schaefer \& Ringwald 1995; Salazar et al. 2017; Schaefer 1990; 2011; and this paper).  For this program we have spent roughly 200 nights on telescopes the world over, as well as roughly 20 trips to plate archives in North America, Europe, and Australia.  Now, with period-changes apparently having some large unknown offset from the simple mass-ejection value, our whole program cannot pay off by measuring the ejected mass.  Further, if this large $\Delta$P is applicable to recurrent novae, then we cannot use the observed large and positive values to demonstrate that their white dwarfs are losing a lot of mass across each eruption, and hence would not be progenitors of Type Ia supernova (Schaefer 2011).  So all this effort on this program can only have other applications.  

The nature of the mechanism for the large $\Delta$P will determine the implications.  Two plausible end-cases are that either the $\Delta$P is the same from eruption-to-eruption for each nova (case 1), or that the $\Delta$P varies randomly over a large positive-to-negative range from eruption-to-eruption (case 2).  For case 1, the large-negative-$\Delta$P for QZ Aur will make every eruption grind down the orbit speedily.  For case 2, a symmetric distribution of positive-and-negative period-changes  means that the orbit jerks longer-and-shorter each eruption, perhaps with the jerks averaging out in the long-run.  Case 2 might be if the direction of the mass imbalance varies widely and randomly from the forward direction of the white dwarf's orbital motion.  The implications for nova evolutionary models changes radically on these two cases.

\section{Comparisons With Models For CV Evolution}

CV evolution is the largest open question for our field.  Our results on QZ Aur can be used to directly test various predictions for CV evolution models.  In the next three subsections, we use our QZ Aur results to evaluate the predictions for three evolution models.

\subsection{Magnetic Braking}

The earliest CV evolution models were based on the angular momentum losses from magnetic braking of the secondary star plus gravitational radiation by the orbiting binary (Rappaport, Joss, \& Webbink 1982, Patterson 1984, Knigge et al. 2011).  Magnetic braking is when the secondary star has a stellar wind that flies out, yet is forced into corotation with the secondary star through the magnetic field, so the ejected wind material carries away rotational angular momentum of the secondary star, while a fast synchronization time quickly turns this into a loss of angular momentum of the orbit.  For systems above the period gap ($P>$3 hours), the magnetic braking dominates and relentlessly drives the orbit to smaller-and-smaller radii.  For systems below the period gap ($P<$2 hours), the magnetic braking largely ceases and so the gravitational radiation dominates the evolution of the system.  For QZ Aur, the magnetic braking should be dominant.

This magnetic braking mechanism is taken to control the very long term evolution of CVs.  After the young CV first comes into contact, the accretion rate is controlled by the magnetic braking, as is the orbital period.  Over time, the braking forces the system to smaller and smaller orbital radius and period, all in a strictly determined manner.  When the orbital period reaches the period gap, the accretion and magnetic braking turn off, leaving the system to grind down slowly by means of the gravitational radiation loss of angular momentum.  When the orbital period has shrunk to below the period gap, the accretion picks back up, and we have the slow evolution of the system down to some minimum orbital period.  This evolution model predicts an inevitable spiral-in of the original binary orbit along a single path, with the nature of the system changing throughout as the orbital period and accretion rate wind down.

Within magnetic braking models, the rate of angular momentum loss ($\dot{P}_{mb}$) is largely determined by just the orbital period (see Equation 26 of Patterson 1984).  For the 8.6 hour orbital period of QZ Aur, magnetic braking models vary somewhat, depending on the assumed magnetic braking law and the particular masses.  From Figure 5 of Rappaport, Verbunt, \& Joss (1983), we see that their middle model has the orbital period fall from 6.9 hours to 6.3 hours over a 10$^7$ year interval, which gives a $\dot{P}_{mb}$ value of -2.4$\times$10$^{-12}$ days per cycle, albeit for a slightly shorter orbital period.  Alternatively, from Figure 11 of Knigge et al. (2011), with a short extrapolation to the period of QZ Aur, they get $\dot{P}_{mb}$ of -3$\times$10$^{-11}$ seconds per second, which translates into -1.1$\times$10$^{-11}$ days/cycle.  QZ Aur has an observed $\dot{P}$ that is the same order-of-magnitude as the predictions of magnetic braking.

For QZ Aur, to estimate the nova recurrence time, $\tau_{rec}$, we can start with the {\it Gaia} distance of 3200$^{+4030}_{-330}$ parsecs (Schaefer 2018), calculate the absolute magnitude in quiescence of +2.7 mag (Schaefer 2018), convert to an approximate accretion rate of $3\times10^{-8}$ M$_{\odot}$/year (from Dubus, Otulakowska-Hypka, \& Lasota 2018), and note the radial velocity measure of the white dwarf mass of 0.93--1.02 M$_{\odot}$ (Szkody \& Ingram 1994).  Then from Townsley \& Bildsten (2005), the ignition mass is 6.3$\times$10$^{-6}$ M$_{\odot}$, for $\tau_{rec}$=210 years.  (By pushing to the closest {\it Gaia} distance, we only get to a recurrence time of 320 years.)  Alternatively, from Yaron et al. (2005), the same input gives $\tau_{rec}$=420 years by interpolation from their big table.  We adopt $\tau_{rec}$=420 years as being the most conservative reasonable value.  This means that every 420 years or so, QZ Aur suffers some additional period-change on top of the magnetic braking.  

The implications or the magnetic braking model depend critically on whether we have case 1 ($\Delta$P remains constant) or case 2 ($\Delta$P varies widely positive-and-negative) from eruption-to-eruption:

For case 1, the constant $\Delta$P from every eruption will grind down the orbital period greatly faster than from magnetic braking.  On a long timescale, the sharp period change each eruption will produce an average rate of period change that can be expressed as a period change averaged over the eruption cycle $\langle \dot{P} \rangle$.  This  is $\Delta P/(\tau_{rec}/P)$, or $\langle \dot{P} \rangle$=-2.4$\times$10$^{-10}$ days per cycle.  This value should be directly compared to $\dot{P}_{mb}$ from magnetic braking.  In units of 10$^{-11}$ days/cycle, the observed $\langle \dot{P} \rangle$ is -24, while the theoretical values for magnetic braking range from -0.24 to -1.1.  That is, the actual time-averaged period change from each eruption is 22$\times$ to 100$\times$ larger than taken by magnetic braking models.  This means that the magnetic braking effect is negligibly small, while some other effect dominates.  It means that all prior evolution-by-magnetic-braking calculations are greatly in error, because they have not included the angular momentum loss through the eruptions.  This means that some other effect is controlling the evolution of CVs.

This presents a severe problem for the magnetic braking model.  It is not that the magnetic braking mechanism does not operate as modeled in many papers, but that magnetic braking is negligibly small, with some other effect dominating the long-term evolution.  It really does not matter what magnetic braking is doing, and the detailed model predictions must fail.  That is, the magnetic braking model is wrong for case 1.

The case 1 situation also greatly changes the expected demographics of CVs.  We can get a timescale for the system lifetime as $\tau_{rec}/(|\Delta P|/P)$, which is 1.4 million years.  (This time scale does not give a useable lifetime measure because $\tau_{rec}$ and $\Delta$P will certainly change greatly over time.)  This will be something like 22--100$\times$ shorter than from the magnetic braking demographics calculations.  With this, all the demographic results from the magnetic braking model will have to be changed and likely overturned.

Before we drop such a venerable model as magnetic braking, we have to explore ways to keep the model alive.  Well, the observed $\Delta$P is robust with no prospect of impeachment.  We could consider that perhaps the derived $\tau_{rec}$ might be 22--100$\times$ larger than given by both Townsley \& Bildsten (2005) and Yaron (2005), but this seems to be greatly too-large a required change to be possible.  (The recent {\it Gaia} DR2 distance puts a lower limit on the distance, which translates to a lower limit on the accretion rate, with the large accretion rate making for building up the trigger mass on the white dwarf in a short time.)  A glib fallback is that maybe there is something special or unique about QZ Aur.  But this fails because QZ Aur does not really have any unusual or extreme property.  The magnetic braking model has a tremendous amount of inertia and calculations behind it, and it has some notable successful `predictions' (primarily the distribution of CV periods), so most workers in our community might be reluctant to discard the venerable model on the basis of our $\Delta$P for QZ Aur alone.  But V1017 Sgr also has a large negative $\Delta$P value which dominates over any magnetic braking, while BT Mon, with a small positive $\Delta$P, has $\dot{P}_{mb}\approx -\langle \dot{P} \rangle$ so that the total effect is near zero.  So three-out-of-three CNe have the magnetic braking being dominated by the period changes across their eruptions.

Case 2 (with randomly positive-and-negative $\Delta$P changing every eruption) changes all this, allowing for the retention of the basic magnetic braking model.  That is, some eruptions will have large-positive-$\Delta$P, others will have small-$\Delta$P, and others will have large-negative-$\Delta$P.  Over evolutionary timescales, covering many individual nova events, the case 2 assumption is that the average $\Delta$P is near zero.  Then, the long term evolution will be described by the magnetic braking case, with jitters up-and-down in the period caused by the eruptions.  The eruption $\Delta$P then just becomes equivalent to a form of noise in the period evolution.  With this case 2 assumption, all the magnetic braking models retain their validity.

A major requirement with case 2 saving magnetic braking models is that the average $\Delta$P must get very close to zero for the braking to dominate.  For example, if the long-term average $\langle \dot{P} \rangle$ is equal in size and opposite in sign to $\dot{P}_{mb}$, then the net effect will be as if there is zero magnetic braking.  If the long-term average $\langle \dot{P} \rangle$ is equal to $\dot{P}_{mb}$ in both size and sign, then it will be as if the magnetic braking will be double that in models, with large changes in the model predictions.  So for the magnetic braking to dominate, we must have $|\langle \dot{P} \rangle|$ be much less than the $|\dot{P}_{mb}|$.  Let us consider a long time interval over which the system drops in $P$ by 0.1 days, a substantial fraction of its evolution as a CV, over which the system will suffer something like $N_{eruptions}=(0.1day)/(\dot{P}_{mb} \tau_{rec})$ eruptions.  For QZ Aur and the above estimates, roughly half-a-million cycles will go round between each eruption, with 20,000 to 90,000 eruptions over the time interval.  For a distribution of period changes across one eruption with an RMS of $\sigma_1$ and a mean of zero, the expected $\langle \dot{P} \rangle$ over the time interval should be roughly $\sigma_1 N_{eruptions}^{-0.5}$.  The best and only estimate is that $\sigma_1\sim \Delta$P/$\tau_{rec}$ as observed.  The long-term average $\langle \dot{P} \rangle$ then has a typical size of 0.08--0.17 in units of 10$^{-11}$ days/cycle.  Thus, we see that for this simple model, we expect the average of the many $\Delta$P values over this long interval to be only somewhat smaller than $\dot{P}_{mb}$.  This is simply because $N_{eruptions}$ is so large over this evolutionary time scale.  Still, it is a close-run issue.  And this still requires that the $\Delta$P mechanism be finely tuned so as to produce an average close to zero.  Indeed, if the average is more than 1--4\% away from zero, then the magnetic braking models are broken.

Given the prior successes of the model (and the social inertia for almost four decades), many in our community will embrace case 2.  Such a case is possible, but it is based on just a speculative assumption, and further that the the $\Delta$P distribution must have an average fine tuning to  $\ll$1--4\%.  Some advocates and modelers of magnetic braking might prefer to ignore the problem by making the critical and unstated speculative assumption with fine-tuning.

\subsection{Hibernation}

The Hibernation model has the strong requirement and prediction that $\Delta$P>0, and a need for $\Delta$P$\gg$0.  This comes from a necessary relation between the observed $\Delta$P and the resulting change in the system brightness.  It starts with Kepler's Law, where a change in the orbital period makes for a change in the semi-major axis.  The change in the Roche lobe size for the secondary is determined from $\Delta$P.  The secondary star cannot change its size on any fast time scale, so the radii of atmospheric pressure levels remains constant across eruptions.  The change in the accretion rate is governed by the change in the Roche lobe size and the atmospheric scale height (see Eq. 4.19 of Frank, King, \& Raine 2002).  The change in the accretion rate is proportional to the change in the accretion disk luminosity, and this luminosity dominates in almost all novae.  So to get Hibernation, with a large drop or cessation of accretion, the Roche lobe size must get substantially larger, which requires that the orbital period get larger by many atmospheric scale heights.  To get a large drop in the system brightness, the orbital period must get larger by a significant amount that can be calculated accurately for each nova.  If $P$ gets smaller across an eruption, then the Roche lobe must get smaller, the accretion rate must increase, and Hibernation cannot happen.  Importantly, this connection from the observed $\Delta$P to the system brightness is completely independent of the mechanisms for period change.  So it does not matter whether the $\Delta$P comes from ejection imbalances or frictional losses of angular momentum in the expanding nova shell or whatever.  If $\Delta$P$<$0, then Hibernation is not working for that eruption.

Our work shows with high confidence that $\Delta$P<0 for QZ Aur, so we know that the Hibernation mechanism is not working.  That is, we have a strong refutation of Hibernation, at least for this one system and this one eruption.

We also have a strong refutation of the Hibernation model for the 1919 nova eruption of V1017 Sgr, where $\Delta$P/P=-273$\pm$61 parts-per-million (Salazar et al. 2017).  Again, the nova's orbital period {\it decreased} by a huge amount, such that the binary separation got smaller, and the secondary star's Roche lobe radius got smaller, and the accretion rate certainly will not become catastrophically lower as part of a Hibernation state.  The measure of $\Delta$P is again of high confidence with no chance of impeachment.  The Hibernation mechanism is certainly not working for the 1919 eruption of V1017 Sgr.

Only BT Mon has a positive-$\Delta$P, and that for its 1939 eruption, with $\Delta$P/P=39$\pm$4.8 ppm (Schaefer \& Patterson 1983).  This is the one case that was part of the original motivation for the Hibernation model back in the 1980s.  Since 1983, BT Mon has been the only known CN with a measured $\Delta$P, thus allowing Hibernation advocates and modelers to think that $\Delta$P>0 is the norm.  However, despite having a positive $\Delta$P, Hibernation is not working for BT Mon.  We know this for two reasons:  The first trouble is that the $\Delta$P is too small.  Detailed calculations (see below) with the change in size of the Roche lobe and the scale height of the secondary star atmosphere points to a change in the accretion rate that is greatly too small to be called `hibernation'.  Second, with our remeasure of the B magnitudes on the Harvard plates, the pre-eruption magnitude outside of eclipse is a steady 15.5 mag or so, while from various sources the post-eruption magnitude from 1962 to present is close to 15.5 mag.  That is, from 23 years to 79 years after the eruption, BT Mon has been holding steady at its pre-eruption magnitude, with no sign of transients from the eruption fading out and no sign of going into a hibernation state.  That is, even BT Mon cannot go into hibernation and is not going into hibernation.

The current situation is that two CNe have $\Delta$P$<$0 and the third CN has $\Delta$P greatly too small to allow for Hibernation.  Hibernation is strongly refuted for 3-out-of-3 eruptions.  On the face of it, Hibernation is dead.

But Hibernation is also a venerable model, widely known almost as long as the magnetic braking models.  So we have to consider ways by which Hibernation can be revived:

A Hibernation advocate might try to separate out QZ Aur as being somehow different, with the Hibernation prediction not being applicable.  But QZ Aur was an ordinary S-class nova with a middle-of-the-pack decline rate, with middling component masses, and with an ordinary $P$.  And QZ Aur ($P$=8.6 hours, M$_{WD}$=0.975$\pm$0.045 M$_{\odot}$, $q$=0.95$\pm$0.05; see Szkody \& Ingram 1994) is strikingly similar to BT Mon ($P$=8.0 hours, M$_{WD}$=1.04$\pm$0.06 M$_{\odot}$, $q$=1.19$\pm$0.06; see Ritter \& Kolb 2003), while they also have similar absolute magnitudes in quiescence (Schaefer 2018), and hence similar accretion rates.  So we cannot point to any of these fundamental properties as being the key for the big decrease in period for QZ Aur.  In all, we cannot pick out QZ Aur as being special or exceptional.

A Hibernation advocate might still wonder about whether QZ Aur is unusual or separate in some way, setting them aside from the Hibernation prediction.  But this fails in light of QZ Aur, V1017 Sgr, and BT Mon spanning the range of the primary nova properties.  These novae represent an ordinary cross section of CNe, sharing no consistent or extreme property.

A Hibernation advocate might try to impeach the data.  But in Salazar et al. (2017) and this paper, we see that the result is robust and confident, and the same is true for the BT Mon measure (Schaefer \& Patterson 1983).  So there is no real chance of any useful claims from this aspect.

We can think of only one possible way out for Hibernation, and that is to adopt a case 2 variation in $\Delta$P.  Case 2 has the random variation of $\Delta$P over a somewhat wide range, centered around zero, so half of the individual nova eruptions have positive-$\Delta$P and at least the possibility that Hibernation can take place.  So half of the eruptions will have no hibernation, while the other half might have hibernation.  There is no fine-tuning problem, as it largely does not matter whether 45\% or 55\% of the eruptions go to hibernation.  The demographics from Hibernation will change substantially, but they are not fitting any close requirements, so this is fine.

We can ask how big the $\Delta$P must be for a hibernation state to be approached.  To get to a state where the accretion is largely turned off, we'd require that the disk have an absolute magnitude below something like +12 mag, while some sort of a stunted low state might be if the accretion disk faded by just 5 mag.  For QZ Aur, we adopt a quiescent absolute magnitude of +2.7 mag (Schaefer 2018), a white dwarf mass of 0.98 M$_{\odot}$, a secondary mass of 0.93 M$_{\odot}$ (Szkody \& Ingram1994), and a secondary surface temperature of 5200 K (Campbell \& Shafter .  The scale height of the secondary atmosphere will be 180 km.  If the orbital period increased by 1700 ppm, then the secondary's Roche lobe would get larger by 830 km, or 4.6 scale-heights.   The ratio of the mass of atmosphere above the Roche lobe will be $e^{-4.6}$ for the two cases, or the accretion rate will drop by close to a factor of 100$\times$, which corresponds to a fading of 5 mag.   To get a drop in the disk luminosity down to an absolute magnitude of +12, we would need a period change by +3200 ppm.  The numbers are similar for BT Mon.  In all, for QZ Aur to go into a low accretion state such that anyone would care to call it `hibernation', the $\Delta$P would have to be both positive and greatly larger in size than we report.  So for QZ Aur to go into a Hibernation state with near-zero accretion, we need $\Delta$P/P$>$+3200 ppm, while a shallow hibernation claim requires $\Delta$P/P$>$+1700 ppm.    

For case 2, with a zero mean, then few if any eruptions will surpass +3200 ppm (or even +1700 ppm), and Hibernation becomes a rare occurrence if ever.  Presumably, a Hibernation advocate can postulate a version of case 2 where the average $\Delta$P/P is very large and positive.  But such would be essentially assuming that which is to be proved.

QZ Aur has a huge period decrease across its 1964 eruption, so the orbit and the Roche lobe necessarily shrunk in size, and the system certainly is not going into hibernation from this eruption.  And 2-out-of-3 CNe have measured $\Delta$P/P$<$0, so Hibernation is certainly not happening in these eruptions.  And 3-out-of-3 are violating the $\Delta$P/P$\gg$+1700 ppm requirement for even shallow Hibernation to work.  The only way that we have thought of to save Hibernation is to make an evidence-less assumption that some substantial fraction of individual eruptions have $\gg$+1700 ppm, despite all three known measures with $\leq$+40 ppm.

\subsection{Asymptotic-Cooling Model}

Patterson et al. (2013) proposed a new model for the evolution of novae, dubbed the `Asymptotic-Cooling' model, making a significant distinction between the paths of novae above and below the period gap.  The asymptotic cooling refers to the diminishing effects of the nova eruption on the system brightness, at first with fast fading, then with an ever slowing rate of fading as the system relaxes to a true state of quiescence.  For systems with periods below the period gap (i.e., $P<$2 hours or so), where gravitational radiation makes for a small accretion rate, each nova system has a very long time between eruptions, so the system has a chance to cool and fade deeply over time, asymptotically approaching a low brightness level, only to have the white dwarf accumulate an ignition mass and go nova again from its low state.  Above the period gap (i.e., $P>$3 hours or so), the magnetic braking drives fast accretion, so the white dwarf accumulates its ignition mass after a relatively short time, while the asymptotic cooling and fading of the system does not have time to get faint.  

Much like for the Hibernation model, as the accretion rates fall and the system fades, the nature of the system changes over to various types of dwarf nova classes.  This model differs from Hibernation in the last half of the cycle, where Hibernation has the system rise in accretion until it becomes high by the time of the next eruption, whereas the Asymptotic-Cooling model cuts out that rising branch and has the nova event come at a time when the accretion rate is at its lowest.

A prediction of the Asymptotic-Cooling model is that the asymptotic brightness level of the system should equal the pre-eruption level, which is to say that the nova erupts from the same brightness level, and that level is what it asymptotically approaches after each eruption.  Necessarily, this makes the further prediction that the long-post-eruption brightness level can never be fainter than the pre-eruption brightness level.  That is, over every eruption cycle, the system relaxes for the same length of time while accumulating the same ignition mass on the white dwarf, so the system is always at the same brightness level at the times of all the nova starts, and this level is always the faintest possible level, so at anytime after a given nova, the tail of the eruption light curve is never fainter than the pre-eruption level.

QZ Aur violates this prediction.  After 1981, the average B-magnitude is 17.44 from 1988--1990 (including the many measures of Campbell \& Shafter) and 17.49 from 2009--2014 (with QZ Aur being far below the pre-eruption level as reported by four observers).  The nature of this drop is not clear (perhaps being a steady decline, or a sudden drop in 1988, or a chaotic set of dips after 1981), but it is clear that, on average, QZ Aur has faded significantly and substantially below the pre-eruption level, by over a factor of two at times.  So the Asymptotic-Cooling model fails for QZ Aur.

There is only one other fully-modern long-term light curve with good pre-eruption levels measured that has appeared in print, and that is for V603 Aql (Johnson et al. 2014).  This is the same nova that is used as the exemplar for the long-period novae in Figure 7 of Patterson et al. (2013).  But V603 Aql has a pre-eruption brightness of B=11.43$\pm$0.03, whereas the post-eruption decline reached that level around two decades after its 1918 eruption, and then continued to fade fairly-steadily down to 12.08$\pm$0.02 in recent years.  This looks to be similar behavior as QZ Aur.  Here is another violator of the model prediction, by nearly a factor of two within some decades after the eruption, and in this case for the exemplar nova.

Again, we must consider dodges to the conclusion that the model prediction has failed for both QZ Aur and V603 Aql.  One reasonable explanation is to note that old novae are always varying on all time scales, so it is fully possible for an old nova out in the asymptotic part of its tail to fluctuate up and down by half a magnitude or more.  In this case, we randomly happened to catch a late downward fluctuation of QZ Aur by a half-magnitude.  We cannot make any far-reaching conclusions while an explanation based on fluctuations is plausible.  This explanation is weak, as we would have to postulate the same happenstance for the two cases where published light curves could reveal such a conundrum.  A second possible explanation is that the inter-eruption time interval need not be identical (perhaps there is hysteresis due to changing temperatures of the white dwarf), so the brightness level at the time of the nova start can fluctuate from cycle to cycle.  In this explanation, QZ Aur is currently asymptoting out to a level of B$\sim$18 for an eruption 400 years from now, but the prior cycle ending in 1964 ended after only 200 years at a level of B=17.16.  But again, this explanation is weak because it requires special circumstances for V603 Aql also, and invoking special pleading for two-out-of-two cases cannot be convincing.

So we are left with QZ Aur and V603 Aql falling significantly and substantially below their pre-eruption levels, with this observed behavior violating the prediction of the Asymptotic-Cooling model.  We can suggest that ordinary variations in old novae (both for the quiescent brightness level and for the inter-eruption times) can make for occasional moderate violations of the prediction that the old novae will not fade fainter than the pre-eruption level.

\section{Conclusions}

QZ Aur was an ordinary S(25) classical nova that erupted in 1964, and it was found in 1995 to have deep eclipses with an 8.6 hour orbital period.  For the primary task of measuring $\Delta$P, the critical and difficult step is to get $P_{pre}$, with this requiring data recorded serendipitously before the 1964 eruption, and the only way to do this is to seek the periodic photometric modulation from eclipses with old archival sky photographs, now stored at observatories around the world.  We have collected a comprehensive photometric history with 1317 magnitudes in a light curve from 1912--2016.  This has included making multiple trips to measure the archival photographic plates at Harvard, Sonneberg, Asiago, and the Vatican, resulting in 60 pre-eruption magnitudes with four separate eclipses.

We have two primary observational results:  Our first result is a confident measure of the orbital period before and after the 1964 eruption.  We find a period of 0.35760096$\pm$0.00000005 days just before, and 0.35749703$\pm$0.00000005 days after, for a {\it decreasing} period, by $\Delta$P/P=-290.71$\pm$0.28 ppm.  This is a robust result, and there is no way around it.

Our second observational result comes from the light curve, with the eclipses and eruption removed.  We find that the light curve is flat and apparently constant with B=17.14 from 1912--1981.  But the later light curve has faded below the pre-eruption level, albeit with significant variability from 17.10--17.98, with averages of B=17.44 in 1988--1990 and B=17.49 in 2009--2014.  This result is strong, and we can prove that all known photometric problems are greatly smaller than is needed to make the drop by 0.35 mag or the variability after the 1980s.

Our two observational results have far-reaching implications, problems, and solutions:

(1) This presents a question to explain how a nova can possibly have its period decrease by the large amount observed across a nova eruption.  Prior work has shown three mechanisms that will change the $\Delta$P value, but even with them all added together we still must have a positive $\Delta$P for the conditions at hand.  The only way to get a huge and negative $\Delta$P is for the nova shell to somehow carry away large amounts of angular momentum.  We point to a new mechanism where asymmetries in the shell ejection can easily carry away enough angular momentum to account for the observed period changes.  Empirically with imaged nova shells, we see that most shells have mass imbalances in the ejecta, with it being common to have $\xi$$>$0.4.  With such an asymmetry, we have a viable mechanism to explain the large period change seen for QZ Aur.

(2) The now-standard evolution model has magnetic braking dominating the long-term evolution of the CVs, where the long-period novae have their fates dictated by this magnetic braking during quiescence.  This is undoubtedly a real effect, and our measured $\dot{P}$ might even come from magnetic braking.  The problem for the model is that the large period decrease during the nova makes for a long-term averaged period change that is 22$\times$ to 100$\times$ larger than the model allows.  With the dominance of the mechanism for angular momentum loss during the eruption, it really does not matter whether or not there is any magnetic braking at all, as this becomes a negligibly small effect.  With this, the substantial successes of the model can only be attributed to the flexibility of the model.  This venerable old model will certainly inspire efforts to reconcile our result for QZ Aur (plus for BT Mon and V1017 Sgr).  We can only think of one dodge to try to save the theory, and that is to make a blind assumption that the $\Delta$P values for a given nova vary eruption-to-eruption over a wide range, both positive and negative, with the average value finely tuned be be close to zero .  In this case (`case 2'), the large jerks in period, both up and down, will average out over the long term, so the remaining effect is just the residual magnetic braking effect.  Such a solution is possible, but it requires an evidence-less assumption and fine-tuning to $\ll$1--4\%, so we take the observed large-negative-$\Delta$P values as a serious challenge to the venerable model of magnetic braking.

(3) The Hibernation model requires $\Delta$P to be positive as the entire mechanism for driving the hibernation.  So if $\Delta$P$<$0, then Hibernation has certainly failed for that eruption.  That $\Delta$P is confidently negative for 2-out-of-3 ordinary CNe is a simple proof that the Hibernation mechanism is {\it not} operating for most novae.  Indeed, for Hibernation to cause a minimal drop by 5.0 mag in the quiescent luminosity of the accretion disk, the standard model requires $\Delta$P/P$\gg$+1700 ppm for both QZ Aur and BT Mon.  So even BT Mon (with its positive $\Delta$P) is not and cannot be going into Hibernation (which agrees with its light curve being closely flat from +10 to +80 years after its eruption), and we are left with 3-out-of-3 CNe with measured $\Delta$P as being far away from any possibility of Hibernation.  This makes for an even-stronger refutation of the Hibernation model.  We see no reasonable possibility for reviving the model in any form.

(4) The Asymptotic-Cooling model has a prediction that the brightness long after the nova event can only be brighter than the pre-eruption level.  Now, QZ Aur has a well measured pre-eruption level of B=17.14, and starting soon after 1981, it has faded to an average level of B=17.49.  So this prediction has failed for QZ Aur (and for V603 Aql).  However, the problem for the model is not severe, because the usual fluctuations in the quiescent brightness as well as plausible variations in the inter-eruption intervals can both readily explain a nova falling below its pre-eruption level by modest levels.

The question of CV evolution is now the broadest and most important for the field.  Our QZ Aur results show that substantial work and revisions are required before a useable model can be presented.

\section*{Acknowledgements}

The AAVSO has provided much that was required for all of the the observers for our program, including finder charts, data archiving, and comparison star magnitudes (through the APASS program).  Funding for APASS has been provided by the Robert Martin Ayers Sciences Fund.  We are grateful for the historical observations and conservation made for the archival plates at the Harvard, Sonneberg, Asiago, and Vatican observatories, by many workers over the last 130 years.  For the particular work in this paper, we are thankful for the hospitality and help from Alison Doane, Josh Grindlay, Peter Kroll, Massimo Turatto, Guy Consolmagno, and Alessandro Omizzolo.  We thank Elizabeth Green for help in taking the light curves at Steward Observatory.  We thank James Arnold for two magnitudes, as reported to the AAVSO data base.


\begin{figure*}
	\includegraphics[width=\columnwidth]{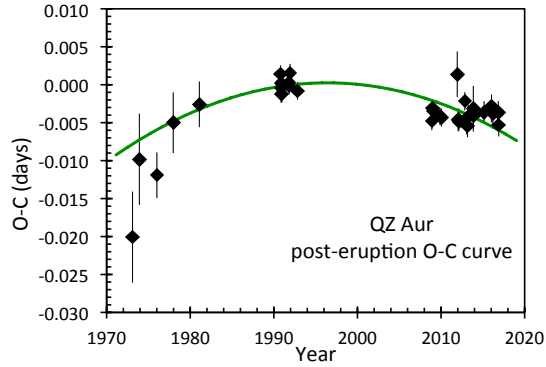}
    \caption{$O-C$ curve for post-eruption eclipse times.  The $O-C$ values are the deviations between the observed eclipse times (from Table 3) and the eclipse times predicted by a linear model (see Eq. 1).  We see that the nova in quiescence certainly has a downward facing curvature (i.e., $\dot{P}$ is negative).  The parabola curve shows the best fit model $O-C$ from the overall joint fit (see Section 6).  (If only the post-eruption eclipse times are fit, then the curvature is slightly larger as the fit seeks to match the eclipse times from around 1975.)  This parabolic model can be extrapolated back to early 1964 at the start of the nova eruption, and this predicted time could provide an accurate epoch that might serve as endpoint on the pre-eruption best fit model.}
\end{figure*}

\begin{figure*}
	\includegraphics[width=\columnwidth]{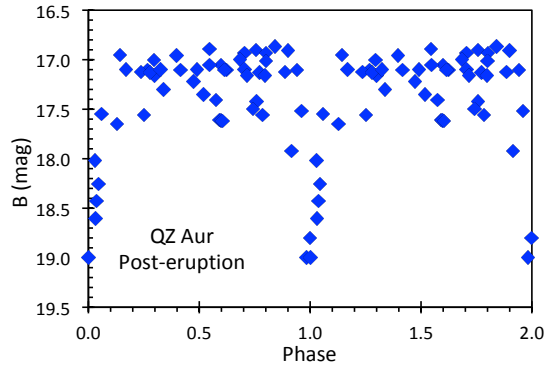}
    \caption{Post-eruption phased light curve.  These 51 post-eruption magnitudes (from Table 1) are almost all from 1967--1990, and are almost all from archival photographic plates.  Each magnitude is plotted twice, once with the phase from 0.0--1.0, and a second time with phase-plus-1.0, to allow the eclipse to be viewed unbroken in the middle of the plot.  This plot is constructed with phases from the best joint fit (see Section 6), although the plot is essentially identical if we use the best fit from the post-eruption eclipse times alone.  The total one-sigma photometric uncertainty is 0.18 mag, as taken from the RMS scatter for the points with phase from 0.1 to 0.9.  What we see is that the many in-eclipse magnitudes are tightly clustered around 1.0, and the many at-maximum magnitudes are spread throughout all phases except for a visible gap during the eclipse.  This is all as it should be, for the confidently known eclipse period.}
\end{figure*}

\begin{figure*}
	\includegraphics[width=\columnwidth]{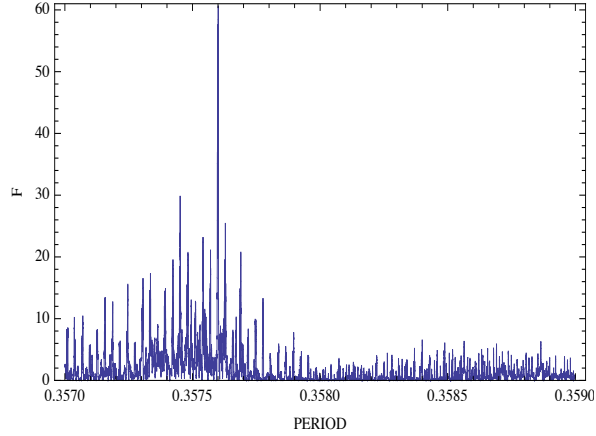}
    \caption{Periodogram for the 60 pre-eruption magnitudes.  We see a high and isolated peak at $P_{pre}$=0.35760076 days, with a peak $F$ value of 60.6.  We also see an approximately-regularly-spaced alias structure (with $F$ from 10 to 29.9) formed when the eclipse times beat with each other.  This search goes from -1400 ppm to +4200 ppm changes, and so all physically plausible values of $\Delta$P/P have been tested.  This plot shows that we have an unique, accurate, and robust measure of $P_{pre}$.}
\end{figure*}

\begin{figure*}
	\includegraphics[width=\columnwidth]{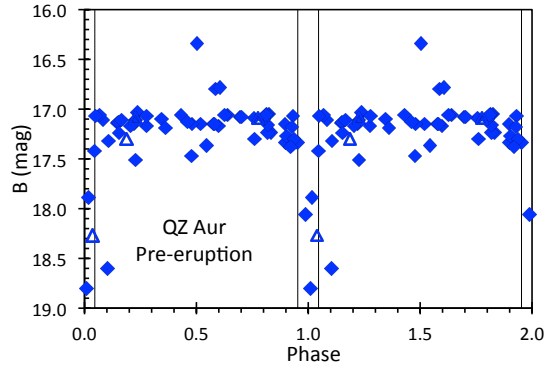}
    \caption{Folded light curve for 60 pre-eruption magnitudes.  The phase for this plot was taken as the values for the joint best fit.  (The best fit with solely pre-eruption data gives essentially the identical plot.)  Four of the magnitudes are upper limits, as represented by triangles.  Each magnitude is plotted twice, once for the phase in the range 0.0--1.0, and a second time for phase+1.0, so that we can well see the eclipse around phase 1.0 with no break.  We see that the eclipse plates are tightly clustered around phase 1.00, while there is a gap in the at-maximum magnitudes over exactly the phase range of the eclipse.  In particular, from phase 0.953--1.046 (as defined by the thin vertical lines), there are zero at-maximum magnitudes and four in-eclipse magnitudes.  With 55 at-maximum magnitudes, we would expect 5 inside the gap if the folding period were wrong, whereas zero are seen.  Visually, we see a classic eclipsing binary light curve, for which the period is confidently identified.}
\end{figure*}

\begin{figure*}
	\includegraphics[width=\columnwidth]{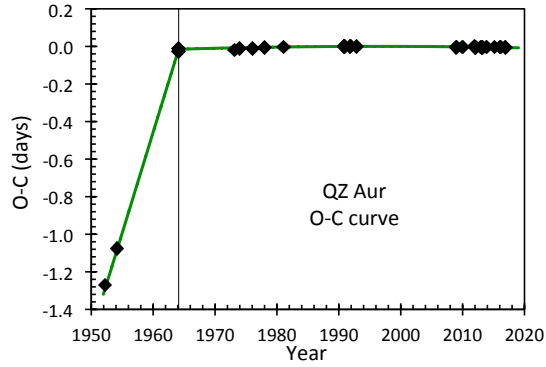}
    \caption{$O-C$ curve for 1952--2016.  This plot is the same as for Figure 1, except that it is extended back in time to include the pre-eruption eclipses from Table 3.  The date of the eruption is indicated with the thin vertical line.  The error bars are all much smaller than the size of each point.  The cycle count for the pre-eruption eclipse times have been confidently determined by the three methods in Section 5.  Further, the cycle count is confirmed by the many pre-eruption plates where they {\it avoid} the line, making a gap in the phased light curve near the zero-phase of mid-eclipse (see Figure 4).  The point of this plot is to show the sharp break, and that the break is downward, so that the $\Delta$P value is large and negative.}
\end{figure*}

\begin{figure*}
	\includegraphics[width=\columnwidth]{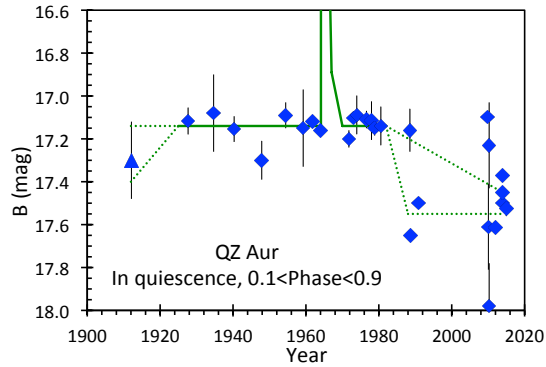}
    \caption{Long-term light curve of QZ Aur in quiescence.  We have selected just the magnitudes out of eclipse and away from eruption, then binned them together.  This gives the behavior in quiescence, which can be compared directly to theoretical predictions.  Away from the 1964 eruption, QZ Aur is consistent with being constant near B=17.14 mag from 1925--1981, with this plus the eruption depicted by the solid curve.  Before 1925, the single upper limit from Harvard (itself with an uncertainty of 0.18 mag) suggests, but does not require, that QZ Aur was fainter in the 1910s.  Around both 1988--1990 and 2009--2014, we have a situation where the brightness appears to be variable from 17.1--18.0 mag.  The uncertain nature of the dimmings below the B=17.14 level are illustrated with dotted lines showing possible cases.  We are confident in the photometry, to within the quoted error bars, so QZ Aur is apparently fast changing.  This can be viewed as the average trend falling below the pre-eruption level.  Or this might be QZ Aur continuing at the pre-eruption level with frequent dips by half-a-magnitude after 1981.  In either case, QZ Aur after 1981 is significantly fainter on average than the pre-eruption level.}
\end{figure*}

\begin{table}
	\centering
	\caption{QZ Aur magnitudes not in long time series}
	\begin{tabular}{lllll} 
		\hline
		HJD & Year & Band & Mag. & Source  \\
		\hline
2419447.6327	&	1912.125	&	B	&	$>$17.3	&	Harvard (MC1649)	\\
2424908.5941	&	1927.076	&	B	&	17.03	&	Harvard (MC22340)	\\
2425183.8268	&	1927.830	&	B	&	17.33	&	Harvard (MC22857)	\\
2425184.8499	&	1927.832	&	B	&	17.30	&	Harvard (MC22863)	\\
2425204.8131	&	1927.887	&	B	&	16.80	&	Harvard (MC22925)	\\
2427313.5622	&	1933.659	&	B	&	16.34	&	Sonneberg (A2001)	\\
2427685.5369	&	1934.678	&	B	&	17.08	&	Sonneberg (A2307)	\\
2429281.5627	&	1939.048	&	B	&	17.23	&	Harvard (MC29952)	\\
2429965.4555	&	1940.919	&	B	&	17.17	&	Sonneberg (GA752)	\\
2429965.5855	&	1940.920	&	B	&	17.06	&	Sonneberg (GA755)	\\
2432504.8799	&	1947.874	&	B	&	17.37	&	Harvard (MC35792)	\\
2432511.7722	&	1947.893	&	B	&	17.23	&	Harvard (MC35806)	\\
2432511.8202	&	1947.893	&	B	&	17.33	&	Harvard (MC35807)	\\
2434100.3056	&	1952.241	&	B	&	18.80	&	Vatican (A1765B)	\\
2434100.3396	&	1952.241	&	B	&	18.60	&	Vatican (A1766B)	\\
2434798.2707	&	1954.151	&	B	&	17.08	&	Vatican (A2553B)	\\
2434798.3057	&	1954.152	&	B	&	17.27	&	Vatican (A2554B)	\\
2434808.3588	&	1954.179	&	B	&	17.89	&	Vatican (A2556)	\\
2434826.2860	&	1954.228	&	B	&	17.14	&	Vatican (A2558)	\\
2435105.8145	&	1954.993	&	B	&	17.05	&	Palomar (252)	\\
2436657.3365	&	1959.242	&	B	&	17.15	&	Sonneberg (A5943)	\\
2437559.5921	&	1961.711	&	B	&	17.17	&	Sonneberg (GC386)	\\
2437560.5051	&	1961.714	&	B	&	17.13	&	Sonneberg (GC390)	\\
2437560.5502	&	1961.714	&	B	&	17.07	&	Sonneberg (GC391)	\\
2437560.6052	&	1961.714	&	B	&	17.06	&	Sonneberg (GC392)	\\
2437561.6053	&	1961.717	&	B	&	17.11	&	Sonneberg (GC398)	\\
2437578.5479	&	1961.763	&	B	&	16.78	&	Sonneberg (GC64)	\\
2437578.6519	&	1961.763	&	B	&	17.15	&	Sonneberg (GC66)	\\
2437582.5422	&	1961.774	&	B	&	$>$17.09	&	Sonneberg (GC80)	\\
2437582.6522	&	1961.774	&	B	&	17.11	&	Sonneberg (GC82)	\\
2437586.6376	&	1961.785	&	B	&	$>$17.09	&	Sonneberg (GC98)	\\
2437588.5158	&	1961.790	&	B	&	17.15	&	Sonneberg (GC114)	\\
2437705.4501	&	1962.111	&	B	&	17.47	&	Sonneberg (GC544)	\\
2437730.3357	&	1962.179	&	B	&	17.06	&	Sonneberg (GC558)	\\
2437731.2977	&	1962.182	&	B	&	17.09	&	Sonneberg (GC560)	\\
2438373.4846	&	1963.941	&	B	&	17.15	&	Sonneberg (GC947)	\\
2438411.2628	&	1964.044	&	B	&	17.15	&	Sonneberg (GC1014)	\\
2438411.4748	&	1964.045	&	B	&	17.05	&	Sonneberg (GC1015)	\\
2438411.5168	&	1964.045	&	B	&	17.07	&	Sonneberg (GC1016)	\\
2438411.5588	&	1964.045	&	B	&	17.07	&	Sonneberg (GC1017)	\\
2438411.6008	&	1964.045	&	B	&	17.11	&	Sonneberg (GC1018)	\\
2438412.2527	&	1964.047	&	B	&	18.06	&	Sonneberg (GC1021)	\\
2438412.2947	&	1964.047	&	B	&	17.32	&	Sonneberg (GC1022)	\\
2438412.3377	&	1964.047	&	B	&	17.51	&	Sonneberg (GC1023)	\\
2438412.3797	&	1964.047	&	B	&	17.10	&	Sonneberg (GC1024)	\\
2438412.4217	&	1964.048	&	B	&	17.13	&	Sonneberg (GC1025)	\\
2438412.4647	&	1964.048	&	B	&	17.15	&	Sonneberg (GC1026)	\\
2438412.5077	&	1964.048	&	B	&	17.08	&	Sonneberg (GC1027)	\\
2438412.5497	&	1964.048	&	B	&	17.16	&	Sonneberg (GC1028)	\\
2438412.5917	&	1964.048	&	B	&	17.30	&	Sonneberg (GC1029)	\\
2438413.2587	&	1964.050	&	B	&	17.10	&	Sonneberg (GC1032)	\\
2438413.3007	&	1964.050	&	B	&	17.38	&	Sonneberg (GC1033)	\\
2438413.3427	&	1964.050	&	B	&	$>$18.27	&	Sonneberg (GC1034)	\\
2438413.3847	&	1964.050	&	B	&	17.24	&	Sonneberg (GC1035)	\\
2438414.3346	&	1964.053	&	B	&	17.06	&	Sonneberg (GC1042)	\\
2438414.3766	&	1964.053	&	B	&	17.18	&	Sonneberg (GC1043)	\\
2438414.4186	&	1964.053	&	B	&	17.42	&	Sonneberg (GC1044)	\\
2438414.4756	&	1964.053	&	B	&	17.17	&	Sonneberg (GC1045)	\\
2438414.5316	&	1964.053	&	B	&	17.19	&	Sonneberg (GC1046)	\\
2438415.3416	&	1964.056	&	B	&	17.06	&	Sonneberg (GC1051)	\\
2439502.4682	&	1967.031	&	B	&	16.89	&	Asiago (549)	\\
2441011.4368	&	1971.163	&	B	&	17.22	&	Asiago (4248)	\\
		\hline
	\end{tabular}
\end{table}

\begin{table}
	\centering
	\contcaption{QZ Aur magnitudes not in long time series}
	\label{tab:continued}
	\begin{tabular}{lllll} 
		\hline
		HJD & Year & Band & Mag. & Source  \\
		\hline

2441055.3354	&	1971.283	&	B	&	17.11	&	Asiago (4294)	\\
2441060.3502	&	1971.297	&	B	&	17.00	&	Asiago (4308)	\\
2441327.5752	&	1972.029	&	B	&	17.56	&	Asiago (5186)	\\
2441391.3471	&	1972.203	&	B	&	17.10	&	Asiago (5277)	\\
2441621.4790	&	1972.832	&	B	&	16.90	&	Asiago (5717)	\\
2441689.3337	&	1973.017	&	V	&	16.41	&	Asiago (6071)	\\
2441693.4613	&	1973.029	&	V	&	17.04	&	Asiago (6103)	\\
2441711.4318	&	1973.078	&	B	&	17.35	&	Asiago (6152)	\\
2441712.2623	&	1973.080	&	B	&	16.87	&	Asiago (6161)	\\
2441718.3660	&	1973.097	&	V	&	17.40	&	Asiago (6214)	\\
2441721.4317	&	1973.105	&	B	&	17.10	&	Asiago (6254)	\\
2441734.3403	&	1973.141	&	B	&	17.05	&	Asiago (6272)	\\
2441737.3948	&	1973.149	&	B	&	16.95	&	Asiago (6292)	\\
2441741.4250	&	1973.160	&	B	&	17.11	&	Asiago (6346)	\\
2441745.4531	&	1973.171	&	B	&	17.00	&	Asiago (6377)	\\
2441763.3062	&	1973.220	&	B	&	17.11	&	Asiago (6423)	\\
2441959.5447	&	1973.758	&	B	&	17.05	&	Asiago (6685)	\\
2442016.5589	&	1973.914	&	V	&	17.50	&	Asiago (6894)	\\
2442121.3798	&	1974.201	&	B	&	17.12	&	Asiago (7135)	\\
2442758.6247	&	1975.947	&	B	&	16.90	&	Vatican (P1679)	\\
2442784.4455	&	1976.018	&	B	&	19.00	&	Asiago (8287)	\\
2442830.4166	&	1976.143	&	B	&	17.41	&	Asiago (8381)	\\
2443100.6203	&	1976.881	&	B	&	16.96	&	Asiago (8752)	\\
2443139.5528	&	1976.988	&	B	&	17.16	&	Asiago (8846)	\\
2443165.4051	&	1977.059	&	B	&	17.11	&	Asiago (8903)	\\
2443492.3946	&	1977.954	&	B	&	17.13	&	Asiago (9376)	\\
2443493.4835	&	1977.957	&	B	&	17.10	&	Asiago (9385)	\\
2443494.4397	&	1977.960	&	B	&	18.80	&	Asiago (9390)	\\
2443543.3959	&	1978.094	&	B	&	17.11	&	Asiago (9454)	\\
2443811.4684	&	1978.829	&	V	&	16.41	&	Asiago (9696)	\\
2443823.6127	&	1978.862	&	B	&	17.13	&	Vatican (P2049)	\\
2443823.6217	&	1978.862	&	B	&	17.16	&	Vatican (P2050)	\\
2443836.4587	&	1978.897	&	B	&	17.10	&	Asiago (9789)	\\
2443836.4775	&	1978.897	&	V	&	16.90	&	Asiago (9790)	\\
2444282.3216	&	1980.118	&	B	&	17.12	&	Asiago (10403)	\\
2444610.4428	&	1981.015	&	B	&	17.16	&	Asiago (10775)	\\
2444637.3572	&	1981.089	&	B	&	19.00	&	Asiago (10868)	\\
2445347.3654	&	1983.034	&	B	&	17.55	&	Asiago (11872)	\\
2446793.7611	&	1986.991	&	B	&	17.52	&	Palomar (SJ01013)	\\
2446828.3808	&	1987.089	&	V	&	16.49	&	Asiago (13706)	\\
2447408.0000	&	1988.675	&	B	&	17.65	&	KPNO (1.3-m)	\\
2447408.0000	&	1988.675	&	V	&	17.18	&	KPNO (1.3-m)	\\
2447408.0000	&	1988.675	&	R	&	16.89	&	KPNO (1.3-m)	\\
2447617.0000	&	1989.245	&	J	&	15.8	&	KPNO (1.3-m)	\\
2447617.0000	&	1989.245	&	K	&	15.5	&	KPNO (1.3-m)	\\
2447891.4096	&	1989.998	&	B	&	17.30	&	Asiago (17988)	\\
2448206.8660	&	1990.861	&	B	&	17.50	&	Mount Laguna (1-m)	\\
2448208.7160	&	1990.866	&	R	&	16.60	&	Mount Laguna (1-m)	\\
2448242.6910	&	1990.959	&	V	&	16.98	&	Mount Laguna (1-m)	\\
2448243.6680	&	1990.961	&	I	&	16.12	&	Mount Laguna (1-m)	\\
2455217.5335	&	2010.054	&	V	&	17.09	&	AAVSO (RMU)	\\
2455291.3569	&	2010.256	&	V	&	16.71	&	AAVSO (RMU)	\\
2455292.3458	&	2010.259	&	V	&	17.46	&	AAVSO (RMU)	\\
2455904.9294	&	2011.936	&	B	&	18.61	&	KPNO (0.8-m)	\\
2455904.9318	&	2011.936	&	B	&	18.43	&	KPNO (0.8-m)	\\
2455904.9342	&	2011.936	&	B	&	18.26	&	KPNO (0.8-m)	\\
2455904.9374	&	2011.936	&	V	&	17.50	&	KPNO (0.8-m)	\\
2455904.9398	&	2011.936	&	V	&	17.41	&	KPNO (0.8-m)	\\
2455904.9422	&	2011.936	&	V	&	17.37	&	KPNO (0.8-m)	\\
2455904.9450	&	2011.936	&	r'	&	17.07	&	KPNO (0.8-m)	\\
2455904.9474	&	2011.936	&	r'	&	17.04	&	KPNO (0.8-m)	\\
2455904.9498	&	2011.936	&	r'	&	17.02	&	KPNO (0.8-m)	\\
		\hline
	\end{tabular}
\end{table}

\begin{table}
	\centering
	\contcaption{QZ Aur magnitudes not in long time series}
	\label{tab:continued}
	\begin{tabular}{lllll} 
		\hline
		HJD & Year & Band & Mag. & Source  \\
		\hline

2455906.9023	&	2011.942	&	r'	&	16.89	&	KPNO (0.8-m)	\\
2455906.9047	&	2011.942	&	r'	&	16.86	&	KPNO (0.8-m)	\\
2455906.9071	&	2011.942	&	r'	&	16.86	&	KPNO (0.8-m)	\\
2455906.9097	&	2011.942	&	V	&	17.07	&	KPNO (0.8-m)	\\
2455906.9121	&	2011.942	&	V	&	17.10	&	KPNO (0.8-m)	\\
2455906.9145	&	2011.942	&	V	&	17.08	&	KPNO (0.8-m)	\\
2455906.9170	&	2011.942	&	B	&	17.61	&	KPNO (0.8-m)	\\
2455906.9194	&	2011.942	&	B	&	17.61	&	KPNO (0.8-m)	\\
2455906.9218	&	2011.942	&	B	&	17.62	&	KPNO (0.8-m)	\\
2457007.5923	&	2014.955	&	V	&	16.96	&	AAVSO (ARJ)	\\
2457009.6268	&	2014.961	&	V	&	17.05	&	AAVSO (ARJ)	\\
		\hline
	\end{tabular}
	
\end{table}

\begin{table}
	\centering
	\caption{QZ Aur time series from 2009 and 2013 (full table is only in the on-line version of this paper}
	\begin{tabular}{lllll} 
		\hline
		HJD & Year & Band & Mag. & Source  \\
		\hline

2455150.9358	&	2009.874	&	V	&	17.19	&	MDM (2.4-m)	\\
2455150.9361	&	2009.874	&	V	&	17.17	&	MDM (2.4-m)	\\
2455150.9363	&	2009.874	&	V	&	17.19	&	MDM (2.4-m)	\\
2455150.9366	&	2009.874	&	V	&	17.20	&	MDM (2.4-m)	\\
2455150.9369	&	2009.874	&	V	&	17.18	&	MDM (2.4-m)	\\
...     &      &        &          &         \\
2456607.0334	&	2013.858	&	B	&	18.49	&	Steward (61-inch)	\\
2456607.0346	&	2013.858	&	B	&	18.64	&	Steward (61-inch)	\\
2456607.0357	&	2013.858	&	B	&	18.72	&	Steward (61-inch)	\\
2456607.0368	&	2013.858	&	B	&	18.78	&	Steward (61-inch)	\\
2456607.0379	&	2013.858	&	B	&	18.83	&	Steward (61-inch)	\\

		\hline
	\end{tabular}
	
\end{table}

\begin{table}
	\centering
	\caption{QZ Aur eclipse times}
	\begin{tabular}{llll} 
		\hline
		$T_i$ & HJD minimum & Year & Source  \\
		\hline

$T_{-4}$	&	2434100.3220	$\pm$	0.008	&	1952.241	&	Vatican (P1765B, P1766B)	\\
$T_{-3}$	&	2434808.3588	$\pm$	0.018	&	1954.179	&	Vatican (P2556)	\\
$T_{-2}$	&	2438412.2527	$\pm$	0.016	&	1964.047	&	Sonneberg (GC1021)	\\
$T_{-1}$	&	2438413.3427	$\pm$	0.015	&	1964.050	&	Sonneberg (GC1034)	\\
$T_{0}$	&	2438440.1496	$\pm$	0.0005	&	1964.122	&	$E_0$	\\
$T_{1}$	&	2441718.3840	$\pm$	0.006	&	1973.097	&	Asiago (6214)	\\
$T_{2}$	&	2442016.5460	$\pm$	0.006	&	1973.914	&	Asiago (6894)	\\
$T_{3}$	&	2442784.4455	$\pm$	0.003	&	1976.018	&	Asiago (8287)	\\
$T_{4}$	&	2443494.4397	$\pm$	0.004	&	1977.960	&	Asiago (9390)	\\
$T_{5}$	&	2444637.3572	$\pm$	0.003	&	1981.089	&	Asiago (10868)	\\
$T_{6}$	&	2448188.0124	$\pm$	0.0012	&	1990.810	&	Campbell and Shafter	\\
$T_{7}$	&	2448205.8846	$\pm$	0.0012	&	1990.859	&	Campbell and Shafter	\\
$T_{8}$	&	2448206.9579	$\pm$	0.0012	&	1990.862	&	Campbell and Shafter	\\
$T_{9}$	&	2448208.7460	$\pm$	0.0012	&	1990.867	&	Campbell and Shafter	\\
$T_{10}$	&	2448242.7080	$\pm$	0.0012	&	1990.960	&	Campbell and Shafter	\\
$T_{11}$	&	2448243.7792	$\pm$	0.0012	&	1990.962	&	Campbell and Shafter	\\
$T_{12}$	&	2448560.8793	$\pm$	0.0012	&	1991.831	&	Campbell and Shafter	\\
$T_{13}$	&	2448561.9522	$\pm$	0.0012	&	1991.834	&	Campbell and Shafter	\\
$T_{14}$	&	2448598.7756	$\pm$	0.0012	&	1991.934	&	Campbell and Shafter	\\
$T_{15}$	&	2448921.9497	$\pm$	0.0012	&	1992.819	&	Campbell and Shafter	\\
$T_{16}$	&	2454799.1816	$\pm$	0.0012	&	2008.909	&	Shi \& Qian	\\
$T_{17}$	&	2454827.0680	$\pm$	0.0012	&	2008.986	&	Shi \& Qian	\\
$T_{18}$	&	2454853.1648	$\pm$	0.0012	&	2009.057	&	Shi \& Qian	\\
$T_{19}$	&	2455150.9584	$\pm$	0.0012	&	2009.874	&	MDM	\\
$T_{20}$	&	2455208.1576	$\pm$	0.0012	&	2010.029	&	Shi \& Qian	\\
$T_{21}$	&	2455904.9232	$\pm$	0.003	&	2011.936	&	KPNO	\\
$T_{22}$	&	2455941.3818	$\pm$	0.0015	&	2011.296	&	AAVSO (BDG)	\\
$T_{23}$	&	2455971.0540	$\pm$	0.0012	&	2012.118	&	Shi \& Qian	\\
$T_{24}$	&	2455973.1987	$\pm$	0.0012	&	2012.124	&	Shi \& Qian	\\
$T_{25}$	&	2456239.1785	$\pm$	0.0012	&	2012.852	&	Shi \& Qian	\\
$T_{26}$	&	2456309.9596	$\pm$	0.0012	&	2013.046	&	Shi \& Qian	\\
$T_{27}$	&	2456355.3615	$\pm$	0.0015	&	2012.857	&	AAVSO (BDG)	\\
$T_{28}$	&	2456385.3924	$\pm$	0.0015	&	2012.970	&	AAVSO (BDG)	\\
$T_{29}$	&	2456605.9676	$\pm$	0.0012	&	2013.855	&	Steward 61"	\\
$T_{30}$	&	2456607.0410	$\pm$	0.003	&	2013.858	&	Steward 61"	\\
$T_{31}$	&	2457075.3604	$\pm$	0.0015	&	2015.572	&	AAVSO (BDG)	\\
$T_{32}$	&	2457395.3196	$\pm$	0.0015	&	2016.778	&	AAVSO (BDG)	\\
$T_{33}$	&	2457402.4702	$\pm$	0.0015	&	2016.805	&	AAVSO (BDG)	\\
$T_{34}$	&	2457461.4560	$\pm$	0.0015	&	2017.027	&	AAVSO (BDG)	\\
$T_{35}$	&	2457698.4761	$\pm$	0.0015	&	2017.921	&	AAVSO (BDG)	\\
$T_{36}$	&	2457721.3542	$\pm$	0.0015	&	2018.007	&	AAVSO (BDG)	\\

		\hline
	\end{tabular}
	
\end{table}

\begin{table}
	\centering
	\caption{QZ Aur best fit ephemerides}
	\begin{tabular}{lrrr} 
		\hline
		   & Post-eruption & Pre-eruption & Joint  \\
		\hline

Observations	&	36 eclipses			&	60 mags			&	36 + 60			\\
$P_{post}$ at $E_0$ (days)	&	0.35749621			&	...			&	0.35749703			\\
	&		$\pm$	0.00000005	&				&		$\pm$	0.00000005	\\
$P_{pre}$ (days)	&	...			&	0.35760093			&	0.35760096			\\
	&				&		$\pm$	0.00000004	&		$\pm$	0.00000005	\\
$E_0$ (HJD)	&	2448555.1591			&	2438440.1498			&	2438440.1497			\\
	&		$\pm$	0.0002	&		$\pm$	0.0007	&		$\pm$	0.0006	\\
$\dot{P}$ (10$^{-11}$ days/cycle)	&	 -2.5$\pm$0.5			&	$\equiv$-3.3			&	 -2.84$\pm$0.22			\\
$\chi^2$ 	&	45.2			&	131.1			&	42.9+131.2			\\
$N_{dof}$	&	33			&	58			&	92			\\
$\Delta$P (days)	&	...			&	...			&	-0.00010393			\\
	&				&				&		$\pm$	0.00000010	\\
$\Delta$P/P (ppm)	&	...			&	...			&	-290.71 	$\pm$	0.28	\\

		\hline
	\end{tabular}
	
\end{table}

\begin{table}
	\centering
	\caption{Light curve for QZ Aur at maximum light}
	\begin{tabular}{lll} 
		\hline
		Year   & B (mag) & Source (Number)  \\
		\hline

1912.12	&	$>$17.3			&	Harvard (1)	\\
1927.66	&	17.12	$\pm$	0.06	&	Harvard (4)	\\
1934.68	&	17.08	$\pm$	0.18	&	Sonneberg (1)	\\
1940.30	&	17.15	$\pm$	0.06	&	Harvard, Sonneberg (3)	\\
1947.88	&	17.30	$\pm$	0.09	&	Harvard (3)	\\
1954.46	&	17.09	$\pm$	0.06	&	Vatican, Palomar (3)	\\
1959.24	&	17.15	$\pm$	0.18	&	Sonneberg (1)	\\
1961.82	&	17.12	$\pm$	0.02	&	Sonneberg (10)	\\
1964.04	&	17.16	$\pm$	0.02	&	Sonneberg (17)	\\
1971.80	&	17.20	$\pm$	0.04	&	Asiago (5)	\\
1973.11	&	17.10	$\pm$	0.02	&	Asiago (10)	\\
1973.96	&	17.09	$\pm$	0.09	&	Asiago (2)	\\
1976.60	&	17.11	$\pm$	0.04	&	Asiago, Vatican (5)	\\
1977.99	&	17.12	$\pm$	0.09	&	Asiago (2)	\\
1978.87	&	17.15	$\pm$	0.04	&	Asiago, Vatican (5)	\\
1980.57	&	17.14	$\pm$	0.09	&	Asiago (2)	\\
1988.54	&	17.16	$\pm$	0.10	&	Asiago (2)	\\
1988.67	&	17.65	$\pm$	0.03	&	Szkody (1)	\\
1990.86	&	17.50	$\pm$	0.02	&	Campbell \& Shafter (many)	\\
2009.87	&	17.10	$\pm$	0.02	&	MDM (8)	\\
2010.05	&	17.61	$\pm$	0.20	&	RMU (1)	\\
2010.26	&	17.23	$\pm$	0.20	&	RMU (1)	\\
2010.26	&	17.98	$\pm$	0.20	&	RMU (1)	\\
2011.94	&	17.61	$\pm$	0.02	&	KPNO (3)	\\
2013.86	&	17.50	$\pm$	0.02	&	Steward (56)	\\
2013.86	&	17.45	$\pm$	0.02	&	Steward (34)	\\
2013.86	&	17.37	$\pm$	0.02	&	Steward (211)	\\
2014.96	&	17.53	$\pm$	0.04	&	ARJ (2)	\\

		\hline
	\end{tabular}
	
\end{table}

\bsp	
\label{lastpage}
\end{document}